\documentclass[aps, nofootinbib, superscriptaddress]{revtex4-2}

\usepackage{amsmath}
\usepackage{amssymb}
\usepackage{graphicx}
\usepackage[utf8]{inputenc}

\usepackage{color}

\newcommand{\hide}[1]{}

\begin{document}

\title{Reciprocal relation of Schwinger pair production between $\textrm{dS}_2$ and $\textrm{AdS}_2$}

\author{Chiang-Mei Chen} \email{cmchen@phy.ncu.edu.tw}
\affiliation{Department of Physics, National Central University, Chungli 32001, Taiwan}
\affiliation{Center for High Energy and High Field Physics (CHiP), National Central University, Chungli 32001, Taiwan}

\author{Chun-Chih Huang} \email{makedate0809@gmail.com}
\affiliation{Department of Physics, National Central University, Chungli 32001, Taiwan}

\author{Sang Pyo Kim} \email{sangkim@kunsan.ac.kr}
\affiliation{Department of Physics, School of Science and Technology, Kunsan National University, Kunsan 54150, Korea}
\affiliation{Center for High Energy and High Field Physics (CHiP), National Central University, Chungli 32001, Taiwan}
\affiliation{Asia Pacific Center for Theoretical Physics, Pohang 37673, Korea}
\affiliation{Center for Relativistic Laser Science, Institute for Basic Science, Gwangju 61005, Korea}

\author{Kuan-Yen Lin} \email{kowal2144@gmail.com}
\affiliation{Department of Physics, National Central University, Chungli 32001, Taiwan}

\date{\today}

\begin{abstract}
The Klein-Gordon and Dirac equation for a massive charged field in a uniform electric field has a symmetry of two-dimensional global de Sitter (dS) and anti-de Sitter (AdS) space. In the in-out formalism the mean numbers of spinors (spin-1/2 fermions) and scalars (spin-0 bosons) spontaneously produced by the uniform electric field are exactly found from the Bogoliubov relations both in the global and planar coordinates of (A)dS$_2$ space. We show that the uniform electric field enhances the production of charged spinor and scalar pairs in the planar and global dS space while the AdS space reduces the pair production in which weak electric fields below the Breitenlohner-Freedman (BF) bound prohibits pair production. The leading Boltzmann factor in dS space can be written as the Gibbons-Hawking radiation or Schwinger effect enhanced by e-folding factors less than one that give the QED effect or the curvature effect. We observe that dS$_2$ and AdS$_2$ spaces are connected by QED, such as a reciprocal relation between the mean number of spinors and scalars provided that the spacetime curvature is analytically continued. The leading behavior of the mean numbers for spinors and scalars is explained as a residue sum of contour integrals of the frequency or momentum in the phase-integral formulation.
\end{abstract}


\maketitle

\section{Introduction}
The $D$-dimensional de Sitter (dS) and anti-de Sitter (AdS) spaces have the symmetry $O(D, 1)$ and $O(D - 1, 2)$ respectively~\cite{Griffiths:2009dfa}, in which the scalar or spinor fields have solutions in terms of special functions~\cite{Chernikov:1968zm, Bousso:2001mw, Avis:1977yn}. The symmetry has been studied in dS space~\cite{Spradlin:2001pw} and AdS space~\cite{Maldacena:2016upp}. The symmetry of (A)dS space gives the group theoretical representation of quantum fields~\cite{Joung:2006gj, Witten:1998qj, Anninos:2019oka}. More interestingly, the extremal limit of nonrotating charged black holes has a near-horizon geometry of AdS space~\cite{Bardeen:1999px} while the Nariai limit of nonrotating charged black holes in dS space has a near-horizon geometry of dS space~\cite{Ortaggio:2002bp}. Near-horizon geometries of extremal black holes are classified, and their symmetries are studied in~\cite{Kunduri:2013gce}.

On the other hand, curved spacetimes or electromagnetic fields can spontaneously produce pairs of particles. Hawking radiation from black holes~\cite{Hawking:1975vcx} and Schwinger pair production by strong electric fields~\cite{Schwinger:1951nm} are two of the most prominent phenomena, which are a nonperturbtive effect of quantum field theory. An expanding spacetime creates particles~\cite{Parker:1968mv}, and particularly a de Sitter space produces pairs of particles from the cosmological horizon~\cite{Gibbons:1977mu}. Reissner-Nordstr\"{o}m black holes emit Hawking radiation of both charge neutral particles and charged pairs with the chemical potential of electrostatic potential~\cite{Gibbons:1975}, and their evaporation of black hole mass and charge has been a continuing interest~\cite{Page:1977um, Hiscock:1990ex, Montero:2019ekk, Brown:2024ajk}.

Extremal charged black holes with two coincident horizons have the zero Hawking temperature but still produce charged pairs due to an electric field on the horizon, the so-called Schwinger pair production. The Schwinger pair production of scalars has been studied in a near-horizon geometry ${\rm AdS}_2 \times {\rm S}^2$~\cite{Chen:2012zn, Chen:2020mqs} while catastrophic emission of charges has been found in the Nariai limit of nonrotating charged black holes in dS space, ${\rm dS}_2 \times {\rm S}^2$~\cite{Chen:2023swn}. Rotating charged black holes have the extremal limit of warped AdS or dS space~\cite{Chen:2021jwy, Chen:2024ctf}. Thus, the spontaneous production of charged pairs by an electric field may shed light on understanding quantum property of charged black holes~\cite{Lin:2024jug}. The Schwinger pair production of scalars has been studied in ${\rm dS}_2$~\cite{Garriga:1994bm, Kim:2008xv} and ${\rm AdS}_2$~\cite{Pioline:2005pf}.

The global geometry of dS space exhibits an interesting aspect that scalar particles are produced in even dimensions but not in odd dimensions~\cite{Mottola:1984ar, Bousso:2001mw}. In fact a massive scalar field equation in the global dS coordinates has reflectionless potentials in odd dimensions, which is a consequence of supersymmetric quantum mechanics~\cite{Cooper:1994eh}. The dimensionality of particle production in dS spaces has been interpreted as the Stokes phenomenon~\cite{Kim:2013iua}, a constructive or destructive interference, in the global coordinates~\cite{Kim:2013cka}. The spinor production in the global dS coordinates also shows the same dimensional behavior~\cite{Jiang:2020evx}. Spinor is also studied in (A)dS space~\cite{Schaub:2023scu}.

Another interesting issue of Schwinger pair production is the boundary condition imposed on quantum fields. In the vector potential for an electric field each positive frequency mode in the past scatters over a barrier and splits into another positive frequency mode and a negative frequency mode in the future, while in the Coulomb potential an incoming (incident) mode scatters by a tunneling barrier into an outgoing (reflected) mode and another outgoing (transmitted) mode~\cite{Nikishov:1970br}. It has been shown in~\cite{Kim:2008yt, Kim:2009pg} that boundary conditions on quantum fields for $E_0/\cosh^2(t/T)$ and $E_0/\cosh^2(x/L)$ result in different formulas for scalars and spinors, which are inverse of each other by the correspondence: $T \leftrightarrow L$, the effective duration and distance, and $\omega \leftrightarrow k$, the energy and momentum. A similar behavior is shown for the emission formulas of scalars from extremal RN and Nariai black holes~\cite{Chen:2020mqs, Chen:2023swn}.

In this paper, we present a comprehensive study of Schwinger pair production for spinors (spin-1/2 fermions) and scalars (spin-0 bosons) induced by a uniform electric field in both the global and planar coordinates of (A)dS$_2$. We also investigate the relationship between the mean numbers of pair production obtained in both coordinates of these two geometries. In the planar coordinates of (A)dS$_2$, Schwinger pair production has been previously studied for scalars~\cite{Garriga:1994bm, Pioline:2005pf, Kim:2008xv, Cai:2014qba} and spinors~\cite{Villalba:1995za, Haouat:2012ik, Stahl:2015cra, Stahl:2015gaa, Botshekananfard:2019zak}. Scalar pair production in the global coordinates was examined in~\cite{Kim:2022nsx}. However, to the best of our knowledge, the Schwinger pair production of spinors in the global coordinates of both dS$_2$ and AdS$_2$ spaces has not yet been investigated.
As the uniform electric field does not break the symmetry of ${\rm (A)dS}_2$ in two dimensions, the Klein-Gordon equation for a massive charged scalar and the Dirac equation for a massive charged spinor can be solved in terms of hypergeometric functions or Whittaker functions, which lead to the exact formulas for Schwinger pairs produced by the electric field which is assisted or suppressed by the scalar curvature. The spinor or scalar production increases in dS space while it decreases in AdS space. The bosonic amplification for scalars is not found, and the Pauli blocking suppresses the mean number less than unity, though more spinors are produced than scalars in AdS space.

Because of the symmetry the mean numbers can be expressed in terms of the Maxwell (invariant) scalar for the electric field and the spacetime curvature, which have the correct limits of the Minkowski space and pure dS space without an electric field, and whose leading Boltzmann factor is given by the ratio of the Maxwell scalar to the scalar curvature square. The exact Schwinger formulas may shed light on understanding how the Maxwell field strongly couples to the spacetime curvature and vice versa.
Remarkably, our results reveal that the mean number of produced scalars or spinors satisfies a reciprocal relation: upon analytic continuation of the curvature, the mean number in dS space becomes the inverse of that in AdS space and vice versa.\footnote{The reciprocality of Schwinger effect between the dS and AdS space in this paper and also the spatial and temporal Sauter-type electric fields should not be confused with the reciprocity relations in thermodynamics, such as the Maxwell and the Onsager-Casimir reciprocity relations~\cite{1945RvMP...17..343C}.}
We apply the phase-integral formulation to explain a physical origin of the leading Boltzmann factor as contributions from simple poles of frequency or momentum in a complex plane of time (space).

The organization of this paper is as follows. In section~\ref{global_dS} we study the Schwinger pair production for scalars and spinors produced by a uniform electric field in the global coordinates of ${\rm dS}_2$. In section~\ref{global_AdS} the mean numbers for scalars and spinors produced by a uniform electric field are found in the global coordinates of ${\rm AdS}_2$. In section~\ref{planar} we find the mean number of spinor production in the planar coordinates of ${\rm (A)dS}_2$. In section~\ref{reciprocal} we discuss the reciprocal relation between the mean numbers of scalars and spinors both in the global ${\rm dS}_2$ and ${\rm AdS}_2$ and in the planar ${\rm dS}_2$ and ${\rm AdS}_2$. In section~\ref{phase_integral} we give a physical interpretation of the Schwinger formulas in the phase-integral formulation. In section~\ref{conclusion} we discuss the physics of the Schwinger production of spinors and scalars.

\section{Pair Production in Global de Sitter Space} \label{global_dS}
In finding explicit formulas for the Schwinger pair production in any dimensional global dS space, an external uniform electric field prohibits both Klein-Gordon and Dirac equations from separating by spherical harmonics, except for dS$_2$ with the scalar curvature $R = 2 H^2$:
\begin{eqnarray} \label{dS2_metric}
ds^2 = - dt^2 + \frac{\cosh^2(H t)}{H^2} d\theta^2, \qquad -\infty < t < \infty, \quad 0 \le \theta \le 2 \pi.
\end{eqnarray}
We then consider a constant electric field with the gauge potential for the global dS$_2$
\begin{eqnarray} \label{dS2_A}
A = - \frac{E \sinh(H t)}{H^2} d\theta \quad \Rightarrow \quad dA = - E \, \vartheta^t \wedge \vartheta^\theta.
\end{eqnarray}
The electric field points the positive (counterclockwise) direction of $\theta$. The Maxwell tensor has components $F_{01} = -F_{10}= -E \cosh(H t)/H$ and gives the Maxwell scalar ${\cal F} = - F_{\mu \nu} F^{\mu \nu}/4 = E^2/2$. The time reversal symmetry of the global metric~(\ref{dS2_metric}) and the anti-symmetry of the gauge field~(\ref{dS2_A}) lead to the time reversal symmetry of the Klein-Gordon or Dirac equation with minimal coupling under the charge conjugation, $q \rightarrow -q$, which is not shared by the planar coordinates, as will be shown explicitly below. It is the two-dimensional $CPT$ invariance of the field equation.

\subsection{Klein-Gordon and Dirac equations}
The Klein-Gordon (KG) equation for a massive charged scalar $\Phi$, carrying mass $m$ and charge $q$, is
\begin{eqnarray} \label{eq_KG}
(D_\mu D^\mu - m^2) \Phi = 0.
\end{eqnarray}
The covariant derivative $D_\mu = \nabla_\mu - i q A_\mu$ includes the spacetime covariant derivative $\nabla_\mu$ and the minimal coupling to gauge field. The equal-time commutator for a positive frequency solution is
\begin{eqnarray}
{\rm Wr} [\Phi(t, x'), \Phi^*(t, x)] = \frac{i}{\sqrt{-g}} \delta(x' - x).
\end{eqnarray}

The Dirac equation for a massive charged spinor field, $\Psi$, is
\begin{eqnarray} \label{eq_Dirac}
\left[ \gamma^a e_a{}^\mu \left( \partial_\mu + \Gamma_\mu - i q A_\mu \right) + m \right] \Psi = 0.
\end{eqnarray}
The tetrad $e_a = e_a{}^\mu \partial_\mu$ and one-form $\Gamma = \Gamma_\mu dx^\mu$ are defined as
\begin{eqnarray}
\eta_{ab} = e_a{}^\mu e_b{}^\nu g_{\mu\nu}, \qquad \Gamma = \Gamma_\mu dx^\mu = \frac18 \left[ \gamma^a, \gamma^b \right] \omega_{ab} = \frac14 \gamma^a \gamma^b \omega_{ab}, \qquad \gamma^a \gamma^b + \gamma^b \gamma^a = 2 \eta^{ab}
\end{eqnarray}
The equal-time anticommutator for the spinor
\begin{eqnarray}
\{ \Psi_{\alpha}(t, x'), \Psi^{\dagger}_{\beta}(t, x) \} = \frac{1}{\sqrt{-g}} \delta(x' - x) \delta_{\alpha \beta},
\end{eqnarray}
after the space integration, in terms of positive $(+)$ and negative $(-)$ frequency modes becomes
\begin{eqnarray}
\Psi^{(+)}_{\alpha}(t) \Psi^{(+)\dagger}_{\beta}(t) + \Psi^{(-)}_{\alpha}(t) \Psi^{(-)\dagger}_{\beta}(t) = \frac{1}{\sqrt{-g}} \delta_{\alpha \beta}.
\end{eqnarray}

\subsection{Charged Scalar Production in dS$_2$}
Separating the angular part $\Phi = \exp(i \ell \theta) \, \phi_\ell(t)$ where $\ell$ is an integer for the periodicity $e^{i \ell \theta} =1$ and setting $\tau = H t$, the KG equation~\eqref{eq_KG} reduces to
\begin{eqnarray}
\left[ \partial_{\tau}^2 + \tanh\tau \, \partial_{\tau} + \frac{2 \ell q E /H^2 \sinh\tau + \ell^2 - q^2 E^2/H^4}{\cosh^2\tau} + \Bigl( \frac{qE}{H^2} \Bigr)^2 + \Bigl( \frac{m}{H} \Bigr)^2 \right] \phi_\ell(\tau) = 0.
 \label{KG_gdS}
\end{eqnarray}
As the global dS metric is time reversal symmetric, the equation is invariant under $\tau \to - \tau$,  together either $\ell \to - \ell$ or $q \to - q$. Once we find a solution, this symmetry can help one get the second solution, for example by $\tau \to -\tau, q \to -q$. The general solution is given by hypergeometric functions:
\begin{eqnarray}
\phi_\ell(\tau) &=& (1 + z)^{\ell/2 + i \kappa/2} (1 - z)^{\ell/2 - i \kappa/2} \left[ C_1 F\left( \frac12 + \ell + i \mu, \frac12 + \ell - i \mu, 1 + \ell + i \kappa; \frac{1 + z}{2} \right) \right.
\nonumber\\
&& \left. + C_2 F\left( \frac12 + \ell - i \mu, \frac12 + \ell + i \mu, 1 + \ell - i \kappa; \frac{1 - z}{2} \right) \right],
\end{eqnarray}
where
\begin{eqnarray}
z = i \sinh \tau, \qquad \kappa = \frac{q E}{H^2}, \qquad \mu = \sqrt{\Bigl( \frac{qE}{H^2} \Bigr)^2 + \Bigl( \frac{m}{H} \Bigr)^2  - \frac{1}{4}}.
\end{eqnarray}

Following Refs.~\cite{1969PhRv..183.1057P} for scalar and~\cite{1971PhRvD...3.2546P} for spinor, the in-state (asymptotically WKB (adiabatic) vacuum) in the past infinity ($\tau = H t \to -\infty$) in a Riemann sheet where
\begin{equation}
\frac{1 + z}2 \to \frac14 \mathrm{e}^{-i \pi/2} \mathrm{e}^{- \tau}, \qquad \frac{1 - z}2 \to \frac14 \mathrm{e}^{i \pi/2} \mathrm{e}^{- \tau},
\end{equation}
we have the following asymptotic behaviors of hypergeometric functions
\begin{eqnarray}
F\left( \frac12 + \ell + i \mu, \frac12 + \ell - i \mu, 1 + \ell + i \kappa; \frac{1 + z}{2} \right) \; &\to& \; A \left( \frac14 \mathrm{e}^{i \pi/2} \mathrm{e}^{- \tau} \right)^{-\frac12 - \ell - i \mu} + B \left( \frac14 \mathrm{e}^{i \pi/2} \mathrm{e}^{- \tau} \right)^{-\frac12 - \ell + i \mu},
\nonumber\\
F\left( \frac12 + \ell - i \mu, \frac12 + \ell + i \mu, 1 + \ell - i \kappa; \frac{1 - z}{2} \right) \; &\to& \; A^* \, \left( \frac14 \mathrm{e}^{-i \pi/2} \mathrm{e}^{- \tau} \right)^{-\frac12 - \ell + i \mu} + B^* \, \left( \frac14 \mathrm{e}^{-i \pi/2} \mathrm{e}^{- \tau} \right)^{-\frac12 - \ell - i \mu},
\end{eqnarray}
where two coefficients are given by
\begin{equation} \label{eq_AB}
A = \frac{\Gamma(1 + \ell + i \kappa) \Gamma(-2 i \mu)}{\Gamma(1/2 + \ell - i \mu) \Gamma(1/2 + i \kappa - i \mu)}, \qquad B = \frac{\Gamma(1 + \ell + i \kappa) \Gamma(2 i \mu)}{\Gamma(1/2 + \ell + i \mu) \Gamma(1/2 + i \kappa + i \mu)}.
\end{equation}
Therefore, the corresponding positive $(+)$ and negative $(-)$ frequency solutions are
\begin{eqnarray}
\phi_\mathrm{in}^{(+)} = \phi_\mathrm{in}^{\rightarrow} &=& 2^{1 + \ell - i 2 \mu} \left[ C_1 B \left( \mathrm{e}^{i \pi/2} \right)^{-1/2 - \ell - i \kappa + i \mu} + C_2 A^* \left( \mathrm{e}^{i \pi/2} \right)^{1/2 + \ell - i \kappa - i \mu} \right] \mathrm{e}^{\tau/2 - i \mu \tau},
\nonumber\\
\phi_\mathrm{in}^{(-)} = \phi_\mathrm{in}^{\leftarrow} &=& 2^{1 + \ell + i 2 \mu} \left[ C_1 A \left( \mathrm{e}^{i \pi/2} \right)^{-1/2 - \ell - i \kappa - i \mu} + C_2 B^* \left( \mathrm{e}^{i \pi/2} \right)^{1/2 + \ell - i \kappa + i \mu} \right] \mathrm{e}^{\tau/2 + i \mu \tau}.
\end{eqnarray}

Similarly, for the out-state in the future infinity ($\tau \to \infty$)
\begin{equation}
\frac{1 + z}2 \to \frac14 \mathrm{e}^{i \pi/2} \mathrm{e}^{\tau}, \qquad \frac{1 - z}2 \to \frac14 \mathrm{e}^{-i \pi/2} \mathrm{e}^{\tau},
\end{equation}
we then have the asymptotic forms
\begin{eqnarray}
F\left( \frac12 + \ell + i \mu, \frac12 + \ell - i \mu, 1 + \ell + i \kappa; \frac{1 + z}{2} \right) \; &\to& \; A \left( \frac14 \mathrm{e}^{-i \pi/2} \mathrm{e}^{\tau} \right)^{-\frac12 - \ell - i \mu} + B \left( \frac14 \mathrm{e}^{-i \pi/2} \mathrm{e}^{\tau} \right)^{-\frac12 - \ell + i \mu},
\nonumber\\
F\left( \frac12 + \ell - i \mu, \frac12 + \ell + i \mu, 1 + \ell - i \kappa; \frac{1 - z}{2} \right) \; &\to& \; A^* \, \left( \frac14 \mathrm{e}^{i \pi/2} \mathrm{e}^{\tau} \right)^{-\frac12 - \ell + i \mu} + B^* \, \left( \frac14 \mathrm{e}^{i \pi/2} \mathrm{e}^{\tau} \right)^{-\frac12 - \ell - i \mu}.
\end{eqnarray}
Consequently, the corresponding positive and negative frequency solutions are
\begin{eqnarray}
\phi_\mathrm{out}^{(+)} = \phi_\mathrm{out}^{\rightarrow} &=& 2^{1 + \ell + i 2 \mu} \left[ C_1 A \left( \mathrm{e}^{i \pi/2} \right)^{1/2 + \ell + i \kappa + i \mu} + C_2 B^* \left( \mathrm{e}^{i \pi/2} \right)^{-1/2 - \ell + i \kappa - i \mu} \right] \mathrm{e}^{-\tau/2 - i \mu \tau},
\nonumber\\
\phi_\mathrm{out}^{(-)} = \phi_\mathrm{out}^{\leftarrow} &=& 2^{1 + \ell - i 2 \mu} \left[ C_1 B \left( \mathrm{e}^{i \pi/2} \right)^{1/2 + \ell + i \kappa - i \mu} + C_2 A^* \left( \mathrm{e}^{i \pi/2} \right)^{-1/2 - \ell + i \kappa + i \mu} \right] \mathrm{e}^{- \tau/2 + i \mu \tau}.
\end{eqnarray}

We impose the boundary condition $\phi_\mathrm{in}^{(-)} = 0$ that describes the scattering of a positive frequency mode of the in-state into a positive and a negative frequency mode of the out-state\footnote{As will be shown in Sec.~\ref{phase_integral}, the canonical time-dependent form~(\ref{can_eq_dS}) for a spinless scalar describes a scattering process with energy $\mu^2$ over a negative potential barrier.}
\begin{eqnarray}
C_2 = - C_1 \frac{A}{B^*} \left( \mathrm{e}^{i \pi/2} \right)^{-1 - 2 \ell - i 2 \mu},
\end{eqnarray}
then the in- and out-states reduce to
\begin{eqnarray}
\phi_\mathrm{in}^{(+)} &=& 2^{1 + \ell - i 2 \mu} \, C_1 B \, \mathrm{e}^{i \pi (1 + 2 \ell)/4} \, \mathrm{e}^{\pi (\kappa - \mu)/2} \Bigl( 1 - \mathrm{e}^{2 \pi \mu} |A|^2/|B|^2 \Bigr) \mathrm{e}^{\tau/2 - i \mu \tau},
\nonumber\\
\phi_\mathrm{out}^{(+)} &=& 2^{1 + \ell + i 2 \mu} \, C_1 A \, \mathrm{e}^{i \pi (1 + 2 \ell)/4} \, \mathrm{e}^{-\pi (\kappa + \mu)/2} \Bigl( 1 + \mathrm{e}^{2 \pi \mu} \Bigr) \mathrm{e}^{-\tau/2 - i \mu \tau},
\\
\phi_\mathrm{out}^{(-)} &=& 2^{1 + \ell - i 2 \mu} \, C_1 B \, \mathrm{e}^{i \pi (1 + 2 \ell)/4} \, \mathrm{e}^{- \pi (\kappa - \mu)/2} \Bigl( 1 + |A|^2/|B|^2 \Bigr) \mathrm{e}^{- \tau/2 + i \mu \tau}.
\nonumber
\end{eqnarray}
Note that the broken time reversal symmetry $\tau \to - \tau$ due to the electric field in~(\ref{KG_gdS}) is restored by the charge conjugation $q \to -q$, which implies that the positive frequency of particle in the future corresponds to that of antiparticle in the past, and vice versa, and whose property is not shared by the planar coordinates since the metric covers different patch under flipping of time coordinate.

For a scattering process, the Bogoliubov coefficients $\alpha$ and $\beta$ defined as~\cite{Chen:2023swn}, with normalized $\phi_\mathrm{in}^{(+)}, \phi_\mathrm{out}^{(+)}$ and $\phi_\mathrm{out}^{(-)}$ without causing confusion of notation
\begin{equation} \label{eq_Bog_def_dS}
\phi_\mathrm{in}^{(+)} = \alpha \, \phi_\mathrm{out}^{(+)} + \beta^* \, \phi_\mathrm{out}^{(-)},
\end{equation}
are given by
\begin{eqnarray}
\alpha =
- 2^{i 4 \mu} \, \frac{\cosh(\pi \mu + \pi \kappa)}{\sinh \pi \mu} \frac{A}{B},
\qquad 
\beta^* =
- \frac{\cosh\pi \kappa}{\sinh\pi \mu},
\end{eqnarray}
in which the coefficients $A$ and $B$ satisfy the following relation
\begin{equation}
\frac{|A|^2}{|B|^2} = \frac{\cosh(\pi \mu - \pi \kappa)}{\cosh(\pi \mu + \pi \kappa)}.
\end{equation}
Then, we obtain the mean number~\cite{Kim:2022nsx}
\begin{eqnarray}
{\cal N}^{\rm (sc)}_{\rm dS}  =  |\beta|^2 = \frac{\cosh^2\pi \kappa}{\sinh^2\pi \mu}.
\label{scalar_ds}
\end{eqnarray}
It should be noted that the mean number~(\ref{scalar_ds}) is invariant under charge conjugation and independent of the mode $\ell$ or its direction. The invariance under charge conjugation is a consequence of the mean number ${\cal N} = \langle \textrm{in} \vert N_{\textrm{out}} \vert \textrm{in} \rangle = \langle \textrm{out} \vert N_{\textrm{in}} \vert \textrm{out} \rangle$ and the equality of produced particles and antiparticles, which are invariant under both the charge conjugation and time reversal of the Klein-Gordon or Dirac equation in the global coordinates. The independence of $\ell$ may be understood from the infinite redshift of physical momentum $\ell_{\rm phys} = \ell /(\cosh (Ht)/H)$ at $t = \pm \infty$, which can be seen by writing the $\tau$-dependent frequency square in~(\ref{KG_gdS}) as $(qE/H^2 \tanh \tau + \ell / \cosh \tau )^2 + (m/H)^2$. Similarly, the particle production from the global coordinates is independent of the harmonic numbers~\cite{Mottola:1984ar}, which is a zero-field limit $E = 0$ of~(\ref{scalar_ds}) in two dimensions.
The vacuum persistence amplitude square
\begin{eqnarray}
|\alpha|^2 = \frac{\cosh(\pi \mu - \pi \kappa) \cosh(\pi \mu + \pi \kappa)}{\sinh^2\pi \mu}
\end{eqnarray}
satisfies the scalar quantization rule
\begin{eqnarray}
|\alpha|^2  - |\beta|^2 = 1.
\end{eqnarray}
Note that the leading Boltzmann factor is ${\cal N}^{\rm (sc)}_{\rm dS} \approx \mathrm{e}^{- 2 \pi (\mu - \kappa)}$ and reduces to the Schwinger formula ${\cal N} = \mathrm{e}^{- \pi m^2/q E}$ in the flat spacetime limit $(H = 0)$. Interestingly, there is no bosonic amplification from the dS space.

\subsection{Charged Spinor Production in dS$_2$}
In order to solve the Dirac equation, we first compute the following related quantities of the global dS$_2$ metric~\eqref{dS2_metric}
\begin{eqnarray}
e_0 = \partial_t, \quad e_1 = \frac{H}{\cosh(H t)} \partial_\theta, \qquad \omega_{01} = - \omega_{10} = - \sinh(H t) \, d\theta, \quad \Gamma_\theta = - \frac12 \sinh(H t) \, \gamma^0 \gamma^1,
\end{eqnarray}
and adopt the Gamma matrices
\begin{eqnarray}
\gamma^0 = \begin{pmatrix} 0 & i \\ i & 0 \end{pmatrix}, \qquad \gamma^1 = \begin{pmatrix} 0 & i \\ -i & 0 \end{pmatrix}.
\end{eqnarray}
The Dirac equation~\eqref{eq_Dirac} for spinor $\Psi = (\psi_\uparrow, \psi_\downarrow)^T$ reduces to
\begin{eqnarray}
\gamma^0 \left( \partial_t + \frac12 H \tanh(H t) \right) \Psi + \frac{H}{\cosh(H t)} \gamma^1 \left( \partial_\theta + \frac{i q E \sinh(H t)}{H^2} \right) \Psi + m \Psi = 0,
\end{eqnarray}
whose components decouple from each other in the second-order formulation 
\begin{eqnarray}
\cosh^2\tau \partial_\tau^2 \psi_\uparrow - \partial_\theta^2 \psi_\uparrow + \sinh\tau \cosh\tau \partial_\tau \psi_\uparrow + ( 1 - 2 i \kappa) \sinh\tau \partial_\theta \psi_\uparrow  + \left[ \left( \mu^2 + \frac14 \right) \sinh^2\tau - i \kappa + \frac{m^2}{H^2} + \frac12 \right] \psi_\uparrow &=& 0,
\nonumber\\
\cosh^2\tau \partial_\tau^2 \psi_\downarrow - \partial_\theta^2 \psi_\downarrow + \sinh\tau \cosh\tau \partial_\tau \psi_\downarrow - ( 1 + 2 i \kappa) \sinh\tau \partial_\theta \psi_\downarrow  + \left[ \left( \mu^2 + \frac14 \right) \sinh^2\tau + i \kappa + \frac{m^2}{H^2} + \frac12 \right] \psi_\downarrow &=& 0,
\end{eqnarray}
where
\begin{equation} \label{kappa_mu_spinor_dS2}
\kappa = \frac{q E}{H^2}, \qquad \mu = \sqrt{\Bigl( \frac{q E}{H^2} \Bigr)^2 + \Bigl( \frac{m}{H} \Bigr)^2}.
\end{equation}

Using the separation of variables $\psi = \exp(i k \theta) \phi(\tau)$ for each component, the general solutions of spinor field $\Psi$ can be written as
\begin{equation}
\Psi = C_1 \begin{pmatrix} \phi_{\uparrow1}(\tau) \\ c_1 \phi_{\downarrow1}(\tau) \end{pmatrix} \mathrm{e}^{i k \theta} + C_2 \begin{pmatrix} \phi_{\uparrow2}(\tau) \\ c_2 \phi_{\downarrow2}(\tau) \end{pmatrix} \mathrm{e}^{i k \theta},
\end{equation}
where $k = \ell + 1/2$ for integer $\ell$ such that the anti-periodicity $\mathrm{e}^{2 i \pi k} = -1$ meets. The coefficients $C_1, C_2$ are two integration constants, and $c_1, c_2$ are ``adjustment'' factors from the coupled equations between the upper and lower components. The general solutions of the upper component are composed of the following independent solutions, in terms of $z = i \sinh\tau$,
\begin{eqnarray} \label{sol_spinor_GdS1}
\phi_{\uparrow1} &=& (1 + z)^{-1/4 + k/2 + i \kappa/2} (1 - z)^{-1/4 - k/2 + i \kappa/2} F\left( i \kappa - i \mu, i \kappa + i \mu, \frac12 - k + i \kappa; \frac{1 - z}{2} \right),
\nonumber\\
\phi_{\uparrow2} &=& (1 + z)^{1/4 - k/2 - i \kappa/2} (1 - z)^{1/4 + k/2 - i \kappa/2} F\left( 1 - i \kappa - i \mu, 1 - i \kappa + i \mu, \frac32 + k - i \kappa; \frac{1 - z}{2} \right),
\end{eqnarray}
and, by the symmetry of equations, the solutions of the lower component can be simply obtained by
\begin{equation} \label{sol_spinor_GdS2}
\phi_{\downarrow1} = \phi_{\uparrow2}(k \to -k, \kappa \to -\kappa), \qquad \phi_{\downarrow2} = \phi_{\uparrow1}(k \to -k, \kappa \to -\kappa),
\end{equation}
and the two additional factors are determined by coupled equations as
\begin{equation}
c_1 = - \frac{m/H}{1 - 2 k + 2 i \kappa}, \qquad c_2 = - \frac{1 + 2 k - 2 i \kappa}{m/H}.
\end{equation}

According to the detailed calculations in Appendix~\ref{App_dS}, for the in-state ($t \to -\infty$) and out-state ($t \to \infty$), the spinor solutions can be expressed in positive/negative frequency modes, with equal normalization of spinor basis, as follows:
\begin{eqnarray}
\Psi_\mathrm{in}^{(+)} \propto \begin{pmatrix} \sqrt{\frac{\mu + \kappa}{\mu}} \\ - \sqrt{\frac{\mu - \kappa}{\mu}} \end{pmatrix} \mathrm{e}^{\tau/2 - i \mu \tau + i k \theta}, &\qquad& \Psi_\mathrm{in}^{(-)} \propto \begin{pmatrix} \sqrt{\frac{\mu - \kappa}{\kappa}} \\ \sqrt{\frac{\kappa + \mu}{\kappa}} \end{pmatrix} \mathrm{e}^{\tau/2 + i \mu \tau + i k \theta},
\nonumber\\
\Psi_\mathrm{out}^{(+)} \propto \begin{pmatrix} \sqrt{\frac{\mu - \kappa}{\mu}} \\ - \sqrt{\frac{\mu + \kappa}{\mu}} \end{pmatrix} \mathrm{e}^{-\tau/2 - i \mu \tau + i k \theta}, &\qquad& \Psi_\mathrm{out}^{(-)} \propto \begin{pmatrix} \sqrt{\frac{\mu + \kappa}{\mu}} \\ \sqrt{\frac{\mu - \kappa}{\mu}} \end{pmatrix} \mathrm{e}^{-\tau/2 + i \mu \tau + i k \theta}.
\end{eqnarray}
Imposing the boundary condition $\Psi_\mathrm{in}^{(-)} = 0$, both the upper and lower components lead to the identical constraint
\begin{equation}
C_2 = - C_1 \frac{B_1}{B_2} 2^{-1 + 2 i \kappa} \mathrm{e}^{(-i - 2 i k - 2 \kappa) \pi/2} = - C_1 \frac{c_1}{c_2} \frac{\tilde B_2'}{\tilde B_1'} 2^{1 + 2 i \kappa} \mathrm{e}^{(i - 2 i k - 2 \kappa) \pi/2}.
\end{equation}
Here, $A$'s and $B$'s are explicitly shown in Appendix A.
Then, the positive frequency in-mode reduces to
\begin{equation}
\Psi_\mathrm{in}^{(+)} = C_1 2^{1/2 + i \kappa - 2 i \mu} \mathrm{e}^{(-i k - \kappa + \mu) \pi/2} \frac{A_1 B_2 - A_2 B_1}{B_2} \sqrt{\frac{\mu}{\mu + \kappa}} \begin{pmatrix} \sqrt{\frac{\mu + \kappa}{\mu}} \\ - \sqrt{\frac{\mu - \kappa}{\mu}} \end{pmatrix} \mathrm{e}^{\tau/2 - i \mu \tau + i k \theta},
\end{equation}
so do the negative and positive frequency out-modes
\begin{eqnarray}
\Psi_\mathrm{out}^{(-)} &=& C_1 2^{1/2 + i \kappa - 2 i \mu} \mathrm{e}^{(i k + \kappa - \mu) \pi/2} \frac{A_1 B_2 - A_2 B_1 \mathrm{e}^{-2 \kappa \pi}}{B_2} \sqrt{\frac{\mu}{\mu + \kappa}} \begin{pmatrix} \sqrt{\frac{\mu + \kappa}{\mu}} \\ \sqrt{\frac{\mu - \kappa}{\mu}} \end{pmatrix} \mathrm{e}^{-\tau/2 + i \mu \tau + i k \theta},
\nonumber\\
\Psi_\mathrm{out}^{(+)} &=& C_1 2^{1/2 + i \kappa + 2 i \mu} \mathrm{e}^{(i k + \kappa + \mu) \pi/2} B_1 (1 - \mathrm{e}^{-2 \kappa \pi}) \sqrt{\frac{\mu}{\mu - \kappa}} \begin{pmatrix} \sqrt{\frac{\mu - \kappa}{\mu}} \\ - \sqrt{\frac{\mu + \kappa}{\mu}} \end{pmatrix} \mathrm{e}^{-\tau/2 - i \mu \tau + i k \theta}.
\end{eqnarray}
By using the relations of $A_1, A_2, B_1, B_2$ (see Appendix~\ref{App_dS}), and then, according to~\eqref{eq_Bog_def_dS}, we find the Bogoliubov coefficients
\begin{eqnarray}
\alpha
= 2^{i 4 \mu} \frac{4 \pi \mu}{\sqrt{\mu^2 - \kappa^2}}  \frac{\Gamma(-2 i \mu)^2}{\Gamma(-i \kappa - i \mu) \Gamma(i \kappa - i \mu) \Gamma(1/2 + k - i \mu) \Gamma(1/2 - k - i \mu)},
\qquad 
\beta^*
= \mathrm{e}^{i k \pi} \frac{\cosh\pi \kappa}{\cosh\pi \mu}.
\end{eqnarray}

The mean number, modulo spin multiplicity, is\footnote{The mean number is a generalization of the $D = 2$ uncharged case in~\cite{Jiang:2020evx}
$$ {\cal N} = \frac{\cos^2(\pi \ell + \pi D/2)}{\cosh^2(\pi m/H)}, $$
with $m/H \to \mu, \, \ell \to \ell + i \kappa$.}
\begin{equation}
{\cal N}^{\rm (sp)}_{\rm dS} = |\beta|^2 = \frac{\cosh^2\pi \kappa}{\cosh^2\pi \mu}.
\label{spinor_ds}
\end{equation}
The vacuum persistence amplitude and mean number satisfy the Bogoliubov relation:
\begin{eqnarray}
|\alpha|^2 = \frac{\sinh (\pi \mu - \pi \kappa) \sinh (\pi \mu + \pi \kappa)}{\cosh^2\pi \mu}, \qquad |\alpha|^2 + |\beta|^2 = 1.
\end{eqnarray}
A physically interesting interpretation of the mean number is the effective temperature~\cite{Cai:2014qba, Chen:2016caa, Volovik:2022cqk}. Two temperatures are associated with the electric field and dS space:
\begin{eqnarray}
T_{\rm U} = \frac{1}{2 \pi} \Bigl(\frac{qE}{m} \Bigr), \qquad T_{\rm GH} = \frac{H}{2 \pi},
\end{eqnarray}
the Unruh temperature for charge acceleration and the Gibbons-Hawking temperature at the cosmological horizon. Then, the mean number has the form
\begin{eqnarray} \label{dS-spinor}
{\cal N}^{\rm (sp)}_{\rm dS} = \mathrm{e}^{ - \frac{m}{T_{\rm GH}} \times \frac{T_{\rm eff}}{T_{\rm GH}} } \times \left( \frac{ \mathrm{e}^{ \frac{m}{T_{\rm GH}} \times \frac{T_{\rm U}}{T_{\rm GH}} }+1}{1 + \mathrm{e}^{- \frac{m}{T_{\rm GH}} \times \frac{\sqrt{T_{\rm U}^2 + T_{\rm GH}^2}}{T_{\rm GH}} } } \right)^2,
\end{eqnarray}
where the effective temperature is
\begin{eqnarray}
T_{\rm eff} = T_{\rm U} + \sqrt{T_{\rm U}^2 + T_{\rm GH}^2}.
\label{eff_tem}
\end{eqnarray}
It should be noted that the Unruh temperature in dS space is $\sqrt{T_{\rm U}^2 + T_{\rm GH}^2}$~\cite{Deser:1998bb} and that the effective temperature~(\ref{eff_tem}) is a characteristic of the Schwinger effect, which becomes $T_{\rm eff} = 2 T_{\rm U}$ in the Minkowski space.
When $m T_{\rm U} \geq 2 \pi T_{\rm GH}^2$, i.e, the dS radius is larger than the radius of the local circular orbit of charge in Euclidean time ($l_{\rm dS} = 1/H \geq l_{\rm E} = 1/\sqrt{qE}$), the mean number is approximated by the leading Boltzmann factor ${\cal N}^{\rm (sp)}_{\rm dS} \simeq \mathrm{e}^{- (m/T_{\rm GH}) \times (\sqrt{T_{\rm U}^2 + T_{\rm GH}^2}-T_{\rm U})/T_{\rm GH}}$. 
In the zero electric field, namely $T_{\rm U} = 0$ and $T_{\rm eff} = T_{\rm GH}$, the mean number becomes Gibbons-Hawking radiation
\begin{eqnarray}
{\cal N} = \left( \frac{2}{\mathrm{e}^{\frac{m}{2T_{\rm GH}}}+ \mathrm{e}^{- \frac{m}{2T_{\rm GH}}}} \right)^2.
\end{eqnarray}
Moreover, when $T_{\rm U} \gg T_{\rm GH}$ ($l_{\rm dS} = 1/H \gg m/qE$), $T_{\rm eff} \simeq 2 T_{\rm U} + T_{\rm GH}^2/2 T_{\rm U}$ and the mean number becomes approximately ${\cal N} \simeq \mathrm{e}^{- m/2 T_{\rm U} }$, the Schwinger formula in the Minkowski space. In the other case of $T_{\rm GH}  \gg T_{\rm U}$, the Boltzmann factor becomes ${\cal N} \simeq \mathrm{e}^{- m/T_{\rm GH} \times (1 -  T_{\rm U}/T_{\rm GH})}$. The emission of charges enhanced by the electric field may be interpreted as Gibbons-Hawking radiation of massive charge with a chemical potential $qE/H = qE l_{\rm dS}$.

Quantitatively, the Boltzmann factor may be written as
\begin{eqnarray}
{\cal N}^{\rm (sp)}_{\rm dS} =
\begin{cases}
\mathrm{e}^{- \frac{m}{T_{\rm GH}} \times \frac{1}{\sqrt{1 + (T_{\rm U}/T_{\rm GH})^2} + T_{\rm U}/T_{\rm GH}}}, \quad &(T_{\rm GH} > T_{\rm U}), \\
\mathrm{e}^{- \frac{m}{2 T_{\rm U}} \times \frac{2}{\sqrt{1+ (T_{\rm GH}/T_{\rm U})^2}+1}}, \quad &(T_{\rm U} > T_{\rm GH}). 
\end{cases}
\end{eqnarray}
The Gibbons-Hawking radiation dominance $(T_{\rm GH} \gg T_{\rm U})$ has an exponential folding factor from the Schwinger or QED effect, and the Schwinger effect dominance $(T_{\rm U} \gg T_{\rm GH})$ has another e-folding factor from the Gibbons-Hawking radiation or spacetime curvature effect. The Gibbons-Hawking radiation and Schwinger effect is enhanced by the QED or curvature effect, respectively, since the e-foldings are always less than one, but the Gibbons-Hawking radiation is more effectively enhanced by QED effect than the Schwinger effect by the curvature.

\section{Pair Production in Global Anti-de Sitter Space} \label{global_AdS}
We now consider the Schwinger effect in the global coordinates of the AdS$_2$ space,
\begin{eqnarray} \label{AdS2_metric}
ds^2 = - \frac{\cosh^2(H \rho)}{H^2} dt^2 + d\rho^2, \qquad 0 \le t \le 2 \pi, \quad 0 < \rho < \infty,
\end{eqnarray}
and a constant electric field with the gauge potential
\begin{eqnarray}
A = \frac{E \sinh(H \rho)}{H^2} dt \quad \Rightarrow \quad dA = - E \, \vartheta^t \wedge \vartheta^\rho.
\end{eqnarray}
Here the electric field is chosen to point the opposite direction to compare with dS space but the final result is independent of direction of the field.

\subsection{Charged Scalar Production in AdS$_2$}
Separating the frequency part $\Phi = \exp(-i \omega t) \, \phi_\omega(\rho)$ and setting $\zeta = H \rho$, the KG equation reduces
\begin{eqnarray}
\left[ \partial_\zeta^2 + \tanh\zeta \, \partial_\zeta + \frac{2 \omega q E/H^2 \sinh\zeta + \omega^2 - q^2 E^2/H^4}{\cosh^2\zeta} + \frac{q^2 E^2}{H^4} - \frac{m^2}{H^2} \right] \phi_\omega(\zeta) = 0.
\end{eqnarray}
The general solution is
\begin{eqnarray}
\phi_\omega(\zeta) &=& (1 + z)^{\omega/2 + i \kappa/2} (1 - z)^{\omega/2 - i \kappa/2} \left[ C_1 F\left( \frac12 + \omega + i \mu, \frac12 + \omega - i \mu, 1 + \omega + i \kappa; \frac{1 + z}{2} \right) \right.
\nonumber\\
&& \left. + C_2 F\left( \frac12 + \omega - i \mu, \frac12 + \omega + i \mu, 1 + \omega - i \kappa; \frac{1 - z}{2} \right) \right],
\end{eqnarray}
where
\begin{eqnarray}
z = i \sinh\zeta, \qquad \kappa = \frac{q E}{H^2}, \qquad \mu = \sqrt{\Bigl( \frac{q E}{H^2} \Bigr)^2 - \Bigl( \frac{m}{H} \Bigr)^2 - \frac{1}{4}}.
\end{eqnarray}

Analogously to the analysis in the dS$_2$ case, the corresponding incoming (incident) and outgoing (reflected) modes for the in-state at $\zeta \to -\infty$ are
\begin{eqnarray}
\phi_\mathrm{in}^{(+)} = \phi_\mathrm{in}^{\rightarrow} &=& 2^{1 + \omega + i 2 \mu} \left[ C_1 A \left( \mathrm{e}^{i \pi/2} \right)^{-1/2 - \omega - i \kappa - i \mu} + C_2 B^* \left( \mathrm{e}^{i \pi/2} \right)^{1/2 + \omega - i \kappa + i \mu} \right] \mathrm{e}^{\zeta/2 + i \mu \zeta},
\nonumber\\
\phi_\mathrm{in}^{(-)} = \phi_\mathrm{in}^{\leftarrow} &=& 2^{1 + \omega - i 2 \mu} \left[ C_1 B \left( \mathrm{e}^{i \pi/2} \right)^{-1/2 - \omega - i \kappa + i \mu} + C_2 A^* \left( \mathrm{e}^{i \pi/2} \right)^{1/2 + \omega - i \kappa - i \mu} \right] \mathrm{e}^{\zeta/2 - i \mu \zeta},
\end{eqnarray}
and for the out-state at $\zeta \to \infty$ the outgoing (transmitted) and incoming modes are
\begin{eqnarray}
\phi_\mathrm{out}^{(+)} = \phi_\mathrm{out}^{\rightarrow} &=& 2^{1 + \omega - i 2 \mu} \left[ C_1 B \left( \mathrm{e}^{i \pi/2} \right)^{1/2 + \omega + i \kappa - i \mu} + C_2 A^* \left( \mathrm{e}^{i \pi/2} \right)^{-1/2 - \omega + i \kappa + i \mu} \right] \mathrm{e}^{-\zeta/2 + i \mu \zeta},
\nonumber\\
\phi_\mathrm{out}^{(-)} = \phi_\mathrm{out}^{\leftarrow} &=& 2^{1 + \omega + i 2 \mu} \left[ C_1 A \left( \mathrm{e}^{i \pi/2} \right)^{1/2 + \omega + i \kappa + i \mu} + C_2 B^* \left( \mathrm{e}^{i \pi/2} \right)^{-1/2 - \omega + i \kappa - i \mu} \right] \mathrm{e}^{-\zeta/2 - i \mu \zeta}.
\end{eqnarray}
Imposing the boundary condition $\phi_\mathrm{out}^{(-)} = 0$ that describes the tunneling process\footnote{As will be shown in Sec.~\ref{phase_integral}, the canonical space-dependent form~(\ref{can_eq_AdS}) for spinless scalar describes a tunneling process with energy $-(m^2/H^2 + 1/4)$ through a positive peaked potential barrier. }
\begin{eqnarray}
C_2 = - C_1 \frac{A}{B^*} \left( \mathrm{e}^{i \pi/2} \right)^{1 + 2 \omega + i 2 \mu},
\end{eqnarray}
the Bogoliubov coefficients $\alpha$ and $\beta$ are defined as~\cite{Chen:2023swn}, with normalized $\phi_\mathrm{in}^{(+)}, \phi_\mathrm{in}^{(-)}$ and $\phi_\mathrm{out}^{(+)}$
\begin{equation} \label{eq_Bog_def_AdS}
\alpha \, \phi_\mathrm{in}^{(+)} + \phi_\mathrm{in}^{(-)} = \beta^* \, \phi_\mathrm{out}^{(+)}.
\end{equation}
It is straightforward to compute
\begin{eqnarray}
\alpha
= 2^{i 4 \mu} \, \frac{\cosh(\pi \kappa + \pi \mu)}{\cosh \pi \kappa} \frac{A}{B},
\qquad 
\beta^*
= \mathrm{e}^{i (1 + 2 \omega)/2} \frac{\sinh\pi \mu}{\cosh\pi \kappa}.
\end{eqnarray}
Finally, we find the mean number~\cite{Kim:2022nsx}
\begin{eqnarray}
{\cal N}^{\rm (sc)}_{\rm AdS} = |\beta|^2 = \frac{\sinh^2\pi \mu}{\cosh^2\pi \kappa},
\end{eqnarray}
and the vacuum persistence amplitude square and the Bogoliubov relation
\begin{eqnarray}
 |\alpha|^2 = \frac{\cosh(\pi \kappa - \pi \mu) \cosh(\pi \kappa + \pi \mu)}{\cosh^2\pi \kappa}, \qquad  |\alpha|^2 -  |\beta|^2 = 1.
\end{eqnarray}

\subsection{Charged Spinor Production in AdS$_2$}
From the related quantities of global AdS$_2$~\eqref{AdS2_metric}
\begin{eqnarray}
e_0 = \frac{H}{\cosh\zeta} \partial_t, \quad e_1 = H \partial_\zeta, \qquad \omega_{01} = - \omega_{10} = - \sinh\zeta \, dt, \quad \Gamma_t = - \frac12 \sinh\zeta \, \gamma^0 \gamma^1,
\end{eqnarray}
the Dirac equation reduces to
\begin{eqnarray}
\frac{1}{\cosh\zeta} \gamma^0 \left( \partial_t - \frac{i q E \sinh\zeta}{H^2} \right) \Psi + \gamma^1 \left( \partial_\zeta + \frac{1}{2} \tanh\zeta \right) \Psi + \frac{m}{H} \Psi = 0.
\end{eqnarray}
The components, with $\Psi = (\psi_\uparrow, \psi_\downarrow)^T$,
have the two second-order formulation
\begin{eqnarray}
\cosh^2\zeta \partial_\zeta^2 \psi_\uparrow - \partial_t^2 \psi_\uparrow + \sinh\zeta \cosh\zeta \partial_\zeta \psi_\uparrow + ( 1 + 2 i \kappa) \sinh\zeta \partial_t \psi_\uparrow  + \left[ \left( \mu^2 + \frac14 \right) \sinh^2\zeta + i \kappa - \frac{m^2}{H^2} + \frac12 \right] \psi_\uparrow &=& 0,
\nonumber\\
\cosh^2\zeta \partial_\zeta^2 \psi_\downarrow - \partial_t^2 \psi_\downarrow + \sinh\zeta \cosh\zeta \partial_\zeta \psi_\downarrow - ( 1 - 2 i \kappa) \sinh\zeta \partial_t \psi_\downarrow  + \left[ \left( \mu^2 + \frac14 \right) \sinh^2\tau - i \kappa - \frac{m^2}{H^2} + \frac12 \right] \psi_\downarrow &=& 0,
\end{eqnarray}
where
\begin{equation} \label{kappa_mu_spinor_AdS2}
\kappa = \frac{q E}{H^2}, \qquad \mu = \sqrt{\Bigl( \frac{q E}{H^2} \Bigr)^2 - \Bigl( \frac{m}{H} \Bigr)^2}.
\end{equation}

Using the separation of variables $\psi = \exp(- i \omega t) \phi(\zeta)$ for each component, the general solutions of spinor field $\Psi$ can be written as
\begin{equation}
\Psi = C_1 \begin{pmatrix} \phi_{\uparrow1}(\zeta) \\ c_1 \phi_{\downarrow1}(\zeta) \end{pmatrix} \mathrm{e}^{-i \omega t} + C_2 \begin{pmatrix} \phi_{\uparrow2}(\zeta) \\ c_2 \phi_{\downarrow2}(\zeta) \end{pmatrix} \mathrm{e}^{-i \omega t},
\end{equation}
where $\omega = \ell + 1/2$ and $\ell$ is integer such that $\exp^{2 i \pi \omega} = -1$. The parameters $C_1, C_2$ are integration constants, and $c_1, c_2$ are constant factors to be fixed by the coupled equations. The general solutions of the upper component are composed of the following independent solutions, in terms of $z = i \sinh\zeta$,
\begin{eqnarray} \label{sol_spinor_GAdS1}
\phi_{\uparrow1} &=& (1 + z)^{-1/4 - \omega/2 - i \kappa/2} (1 - z)^{-1/4 + \omega/2 - i \kappa/2} F\left( -i \kappa - i \mu, -i \kappa + i \mu, \frac12 + \omega - i \kappa; \frac{1 - z}{2} \right),
\nonumber\\
\phi_{\uparrow2} &=& (1 + z)^{1/4 + \omega/2 + i \kappa/2} (1 - z)^{1/4 - \omega/2 + i \kappa/2} F\left( 1 + i \kappa - i \mu, 1 + i \kappa + i \mu, \frac32 - \omega + i \kappa; \frac{1 - z}{2} \right),
\end{eqnarray}
and, by the symmetry of equations, the solutions of the lower component can be obtained by a simple flipping of parameters
\begin{equation}
\phi_{\downarrow1} = \phi_{\uparrow2}(\omega \to -\omega, \kappa \to - \kappa), \qquad \phi_{\downarrow2} = \phi_{\uparrow1}(\omega \to -\omega, \kappa \to - \kappa),
\end{equation}
and the two factors are given by
\begin{equation}
c_1 = - \frac{m/H}{1 + 2 \omega - 2 i \kappa}, \qquad c_2 = \frac{1 - 2 \omega + 2 i \kappa}{m/H}.
\end{equation}

According to the detailed calculations in Appendix~\ref{App_AdS}, the spinor solutions for the in-mode ($\zeta \to -\infty$) and out-mode ($\zeta \to \infty$) have the asymptotic forms
\begin{eqnarray}
\Psi_\mathrm{in}^{(+)} \propto \begin{pmatrix} \sqrt{\frac{\kappa + \mu}{\mu}} \\ - \sqrt{\frac{\kappa - \mu}{\mu}} \end{pmatrix} \mathrm{e}^{\zeta/2 - i \omega \tau + i \mu \zeta}, &\qquad& \Psi_\mathrm{in}^{(-)} \propto \begin{pmatrix} \sqrt{\frac{\kappa - \mu}{\kappa}} \\ \sqrt{\frac{\kappa + \mu}{\kappa}} \end{pmatrix} \mathrm{e}^{\zeta/2 - i \omega \tau - i \mu \zeta},
\nonumber\\
\Psi_\mathrm{out}^{(+)} \propto \begin{pmatrix} \sqrt{\frac{\kappa - \mu}{\mu}} \\ - \sqrt{\frac{\kappa + \mu}{\mu}} \end{pmatrix} \mathrm{e}^{-\zeta/2 - i \omega \tau + i \mu \zeta}, &\qquad& \Psi_\mathrm{out}^{(-)} \propto \begin{pmatrix} \sqrt{\frac{\kappa + \mu}{\mu}} \\ \sqrt{\frac{\kappa - \mu}{\mu}} \end{pmatrix} \mathrm{e}^{-\zeta/2 - i \omega \tau - i \mu \zeta}.
\end{eqnarray}
The tunneling boundary condition $\Psi_\mathrm{out}^{(-)} = 0$, for both the upper and lower components, leads to
\begin{equation}
C_2 = - C_1 \frac{B_1}{B_2} 2^{-1 - 2 i \kappa} \mathrm{e}^{(i - 2 i \omega - 2 \kappa) \pi/2} = - C_1 \frac{c_1}{c_2} \frac{\tilde B_2'}{\tilde B_1'} 2^{1 - 2 i \kappa} \mathrm{e}^{(-i - 2 i \omega - 2 \kappa) \pi/2}.
\end{equation}
Here $A$'s and $B$'s are in the Appendix B. The remaining outgoing (transmitted)  mode (out-mode) is
\begin{equation}
\Psi_\mathrm{out}^{(+)} = C_1^+ 2^{1/2 - i \kappa - 2 i \mu} \mathrm{e}^{(-i \omega - \kappa - \mu) \pi/2} \frac{A_1 B_2 - A_2 B_1}{B_2} \sqrt{\frac{\kappa}{\kappa - \mu}} \begin{pmatrix} \sqrt{\frac{\kappa - \mu}{\kappa}} \\ - \sqrt{\frac{\kappa + \mu}{\kappa}} \end{pmatrix} \mathrm{e}^{-\zeta/2 - i \omega t + i \mu \zeta},
\end{equation}
while the incoming (incident) mode and the outgoing (reflected) mode, all the in-modes, become
\begin{eqnarray}
\Psi_\mathrm{in}^{(-)} &=& C_1^+ 2^{1/2 - i \kappa - 2 i \mu} \mathrm{e}^{(i \omega + \kappa + \mu) \pi/2} \frac{A_1 B_2 - A_2 B_1 \mathrm{e}^{-2 \kappa \pi}}{B_2} \sqrt{\frac{\kappa}{\kappa - \mu}} \begin{pmatrix} \sqrt{\frac{\kappa - \mu}{\kappa}} \\ \sqrt{\frac{\kappa + \mu}{\kappa}} \end{pmatrix} \mathrm{e}^{\zeta/2 - i \omega t - i \mu \zeta},
\nonumber\\
\Psi_\mathrm{in}^{(+)} &=& C_1^+ 2^{1/2 - i \kappa + 2 i \mu} \mathrm{e}^{(i \omega + \kappa - \mu) \pi/2} B_1 (1 - \mathrm{e}^{-2 \kappa \pi}) \sqrt{\frac{\kappa}{\kappa + \mu}} \begin{pmatrix} \sqrt{\frac{\kappa + \mu}{\kappa}} \\ - \sqrt{\frac{\kappa - \mu}{\kappa}} \end{pmatrix} \mathrm{e}^{\zeta/2 - i \omega t + i \mu \zeta}.
\end{eqnarray}

By using the relations of $A_1, A_2, B_1, B_2$ (see Appendix~\ref{App_AdS}), and then from~\eqref{eq_Bog_def_AdS}, we obtain the Bogoliubov coefficients
\begin{eqnarray}
\alpha
&=& 2^{i 4 \mu} \mathrm{e}^{-i \omega \pi} \frac{4 \pi \mu}{\sqrt{\mu^2 - \kappa^2}} \frac{\cosh\pi \mu}{\cosh\pi \kappa} \frac{\Gamma(-2 i \mu)^2}{\Gamma(-i \kappa - i \mu) \Gamma(i \kappa - i \mu) \Gamma(1/2 + \omega - i \mu) \Gamma(1/2 - \omega - i \mu)},
\nonumber\\
\beta^*
&=& \mathrm{e}^{-i \omega \pi} \frac{\cosh\pi \mu}{\cosh\pi \kappa}.
\end{eqnarray}
The mean number and the vacuum persistence amplitude, modulo spin multiplicity, are
\begin{eqnarray}
{\cal N}^{\rm (sp)}_{\rm AdS} = |\beta|^2 = \frac{\cosh^2\pi \mu}{\cosh^2\pi \kappa}, \qquad |\alpha|^2 = \frac{\sinh(\pi \kappa - \pi \mu) \sinh(\pi \kappa + \pi \mu)}{\cosh^2\pi \kappa}.
\end{eqnarray}
In terms of the temperatures, $T_{\rm U} = q E/2 \pi m, T_{\rm GH} = H/2 \pi$, the mean number has the form
\begin{eqnarray}
{\cal N}^{\rm (sp)}_{\rm AdS} = \mathrm{e}^{- \frac{m}{T_{\rm GH}} \times \frac{T_{\rm eff}}{T_{\rm GH}} } \times \left( \frac{1 + \mathrm{e}^{\frac{m}{T_{\rm GH}} \times \frac{\sqrt{T_{\rm U}^2 - T_{\rm GH}^2}}{T_{\rm GH}}} }{1 + \mathrm{e}^{- \frac{m}{T_{\rm GH}} \times \frac{T_{\rm U}}{T_{\rm GH} } }} \right)^2,
\end{eqnarray}
where the effective temperature is
\begin{eqnarray}
T_{\rm eff} = T_{\rm U} + \sqrt{T_{\rm U}^2 - T_{\rm GH}^2}.
\end{eqnarray}
The BF bound for the stability of AdS space is given by $T_{\rm U} \leq T_{\rm GH}$, i.e., $|qE| \leq mH$. A physical meaning is that the circular motion of charge $q$ by the electric field in the Euclidean time extends over the AdS space. Pair production requires circular loops to reside inside the AdS radius, and thus the Schwinger pair production violates the BF bound. The e-folding factor for the Unruh effect is greater than one and suppresses the Schwinger effect.

\section{Charged Spinor Production in Planar (A)dS Coordinates} \label{planar}
In order to compare the Schwinger effect in the global coordinates with that in the planar coordinates, we revisit Schwinger pair production of spinors in the planar coordinates of ${\rm dS}_2$ and, for the first time, study Schwinger pair production of spinors in the planar coordinates of ${\rm AdS}_2$. Then, we will show the relation of the Schwinger pair production between the planar ${\rm dS}_2$ and  ${\rm AdS}_2$. 

The planar ${\rm dS}_2$ coordinates cover a half of the hyperboloid in three-dimensional Minkowski space and have the metric
\begin{eqnarray}
ds^2 = - dt^2 + \mathrm{e}^{2 H t} dx^2, \qquad - \infty < t, \, x < \infty.
\label{pl_dS}
\end{eqnarray}
The uniform electric field has a one-form potential
\begin{eqnarray} \label{pl-pot1}
A = - \frac{E}{H} \bigl( \mathrm{e}^{Ht} - 1 \bigr) dx.
\end{eqnarray}
Under time reversal $t \rightarrow -t$, the metric (\ref{pl_dS}) maps to the other patch of the global Penrose diagram~\cite{Spradlin:2001pw}. 
The planar coordinates for the ${\rm AdS}_2$ space
\begin{eqnarray}
ds^2 = - \mathrm{e}^{2 H x} dt^2 + dx^2, \qquad - \infty < t, \, x < \infty,
\end{eqnarray}
give another one-form potential for the uniform electric field
\begin{eqnarray} \label{pl-pot2}
A = \frac{E}{H} \bigl( \mathrm{e}^{H x} - 1 \bigr)dt.
\end{eqnarray}
When $H \rightarrow 0$, the metric tensors and potentials have the limit of the Minkowski space, i.e. $A = - E t dx$ and $A = E x dt$, respectively.

\subsection{Spinor Production in dS$_2$}
The Dirac equation for spinor field $\Psi = ( \psi_\uparrow, \psi_\downarrow )^T$ in the potential~(\ref{pl-pot1})
\begin{eqnarray}
i \left[ \partial_t + \mathrm{e}^{- H t} \bigl( \partial_x - i q A_x \bigr) + \frac{H}{2} \right] \psi_\downarrow + m \psi_\uparrow &=& 0,
\nonumber\\
i \left[ \partial_t - \mathrm{e}^{- H t} \bigl( \partial_x - i q A_x \bigr) + \frac{H}{2} \right] \psi_\uparrow + m \psi_\downarrow &=& 0,
\end{eqnarray}
leads to the second-order formulation for the components $\psi_a(t, x) = \exp(i k x) \phi_a(t)$
\begin{equation}
\left[ \partial_\tau^2 + \partial_\tau + \mu^2 + \frac14 + 2 \bar{k} \Bigl( \kappa + \frac{i}{2} \Bigr) \mathrm{e}^{- \tau} + \bar{k}^2 \mathrm{e}^{-2 \tau} \right] \phi_\uparrow(\tau) = 0,
\end{equation}
where parameters $\kappa$ and $\mu$ are defined in~\eqref{kappa_mu_spinor_dS2} with $\tau = H t$ and $\bar{k} = k/H - \kappa$, and the other component
\begin{equation}
\left[ \partial_\tau^2 + \partial_\tau + \mu^2 + \frac14 + 2 \bar{k} \Bigl( \kappa - \frac{i}{2} \Bigr) \mathrm{e}^{- \tau} + \bar{k}^2 \mathrm{e}^{-2 \tau} \right] \phi_\downarrow(\tau) = 0.
\end{equation}
The general solutions, in terms of $z = - 2 i \bar{k} \exp(-\tau) $, are
\begin{equation}
\Psi(t, x) = C_1 \begin{pmatrix} M_{-1/2 + i \kappa, i \mu}(z) \\ c_1 M_{1/2 + i \kappa, i \mu}(z) \end{pmatrix} \mathrm{e}^{i k x} + C_2 \begin{pmatrix} W_{-1/2 + i \kappa, i \mu}(z) \\ c_2 W_{1/2 + i \kappa, i \mu}(z) \end{pmatrix} \mathrm{e}^{i k x}
\end{equation}
with
\begin{equation}
c_1 = - \frac{H}{m} (\kappa + \mu), \qquad c_2 = - i \frac{H}{m}.
\end{equation}
Here, $M$ and $W$ denote the Whittaker functions. Unlike in the global coordinates, here the flipping of the wave vector $k \to - k$ can not be recovered by charge conjugation $q \to - q$. Therefore the production rate is different for positive and negative values of $k$.

After decomposing the positive and negative frequency solutions in the past infinity ($\tau \rightarrow - \infty, |z| \rightarrow \infty $) and in the future infinity ($\tau \rightarrow \infty, |z| \rightarrow 0$), we can straightforwardly compute the Bogoliubov coefficients. Finally, the mean number of produced spinor pairs, modulo spin multiplicity, is~\cite{Frob:2014zka, Stahl:2015gaa}
\begin{equation}
{\cal N}_{\rm dS}^\mathrm{(sp)} = \frac{\sinh(\pi \mu + \mathrm{sgn}(\bar{k}) \, \pi \kappa)}{\mathrm{e}^{\pi \mu - \mathrm{sgn}(\bar{k}) \, \pi \kappa} \sinh(2 \pi \mu)}, \qquad \mathrm{sgn}(\bar{k}) = \frac{\bar{k}}{|\bar{k}|}.
\end{equation}
A passing remark is that the mean number of pair production depends on $\mathrm{sgn}(\bar{k})$, which corresponds to the screening and antiscreening of the electric field  by produced pairs in the planar coordinates~\cite{Garriga:1993fh, Frob:2014zka}. Pairs are more produced in the screening case than the antiscreening case, as seen from the Boltzmann factor ${\cal N}_{\rm dS} \simeq \mathrm{e}^{-2 \pi (\mu - \mathrm{sgn}(\bar{k}) \, \kappa)}$.

\subsection{Spinor Production in AdS$_2$}
The two components with $\psi_a(t, x) = \exp(- i \omega t) \phi_a (x)$ of the Dirac equation in the potential~(\ref{pl-pot2}) have the second-order formulation
\begin{eqnarray}
\left[ \partial_\zeta^2 + \partial_\zeta + \mu^2 + \frac14  + 2 \bar{\omega} \Bigl( \kappa - \frac{i}{2} \Bigr) \mathrm{e}^{- \zeta} + \bar{\omega}^2 \mathrm{e}^{-2 \zeta} \right] \phi_\uparrow(\zeta) &=& 0,
\nonumber\\
\left[ \partial_\zeta^2 + \partial_\zeta + \mu^2 + \frac14 + 2 \bar{\omega} \Bigl( \kappa + \frac{i}{2} \Bigr) \mathrm{e}^{- \zeta} + \bar{\omega}^2 \mathrm{e}^{-2 \zeta} \right] \phi_\downarrow(\zeta) &=& 0,
\end{eqnarray}
where parameters $\kappa$ and $\mu$ are defined in~\eqref{kappa_mu_spinor_AdS2} with $\zeta = H x$ and $\bar{\omega} = \omega/H - \kappa$.

The general solutions, in terms of $z = - 2 i \bar{\omega} \exp(-\zeta)$, are
\begin{equation}
\Psi(t, x) = C_1 \begin{pmatrix} M_{1/2 + i \kappa, i \mu}(z) \\ c_1 M_{-1/2 + i \kappa, i \mu}(z) \end{pmatrix} \mathrm{e}^{-i \omega t} + C_2 \begin{pmatrix} W_{1/2 + i \kappa, i \mu}(z) \\ c_2 W_{-1/2 + i \kappa, i \mu}(z) \end{pmatrix} \mathrm{e}^{-i \omega t}
\end{equation}
with
\begin{equation}
c_1 = - \frac{m}{H} \frac1{\kappa + \mu}, \qquad c_2 = i \frac{m}{H}.
\end{equation}
Then we obtain the mean number of produced spinor pairs, modulo spin multiplicity,
\begin{equation}
{\cal N}_{\rm AdS}^\mathrm{(sc)} = \frac{\mathrm{e}^{\pi \mu - \mathrm{sgn}(\bar{\omega}) \, \pi \kappa} \sinh(2 \pi \mu)}{\sinh(\pi \mu + \mathrm{sgn}(\bar{\omega}) \, \pi \kappa)}, \qquad \mathrm{sgn}(\bar{\omega}) = \frac{\bar{\omega}}{|\bar{\omega}|}.
\end{equation}

\section{Reciprocal Relation between dS$_2$ and AdS$_2$} \label{reciprocal}
We observe that the mean numbers of spinors in ${\rm dS}_2$ and ${\rm AdS}_2$ are inverse of each other provided that the scalar curvature is analytically continued between two spaces. In terms of the scalar curvature $R = \pm 2 H^2$ (lower sign for AdS), we express the parameters in terms of the curvature
\begin{eqnarray}
\mu(R) = \sqrt{\Bigl( \frac{2 q E}{R} \Bigr)^2 + \frac{2 m^2}{R}}, \qquad \kappa(R) = \sqrt{ \left( \frac{2 q E}{R} \right)^2}.
\end{eqnarray}
In the global coordinates, the mean numbers of spinor production
\begin{eqnarray}
{\cal N}_{\rm dS}^{\rm (sp)} = \left( \frac{\cosh\pi \kappa(R_{\rm dS})}{\cosh\pi \mu(R_{\rm dS})} \right)^2, \qquad {\cal N}_{\rm AdS}^{\rm (sp)} = \left( \frac{\cosh\pi \mu(R_{\rm AdS})}{\cosh\pi \kappa(R_{\rm AdS})} \right)^2,
\end{eqnarray}
satisfy the reciprocal relation
\begin{eqnarray}
{\cal N}_{\rm dS}^{\rm (sp)}(R) \, {\cal N}_{\rm AdS}^{\rm (sp)}(R) = 1, \label{rec_rel}
\end{eqnarray}
with $R$ denoting the curvature either for ${\rm dS}_2$ or ${\rm AdS}_2$. This means that the mean number ${\cal N}_{\rm AdS}^{\rm (sp)} (R_{\rm AdS})$ in ${\rm AdS}_2$ equals to $1/{\cal N}_{\rm dS}^{\rm (sp)} (R_{\rm AdS})$, and vice versa. However, for fixed value $qE$ and $H$, the mean numbers satisfy the inequality ${\cal N}_{\rm dS}^{\rm (sp)} {\cal N}_{\rm AdS}^{\rm (sp)} \leq 1$, since $R_{\rm AdS} = - R_{\rm dS} < 0$, in which the equality holds when $qE \gg mH$, the Schwinger effect dominance. Similarly, the reciprocal relation holds in the planar coordinates.

It is interesting to note the reciprocal relation between the mean numbers of scalars in the global ${\rm (A)dS}_2$~\cite{Kim:2022nsx}
\begin{eqnarray}
{\cal N}_{\rm dS}^{\rm (sc)} = \left( \frac{\cosh\pi \kappa(R_{\rm dS})}{\sinh\pi \bar{\mu}(R_{\rm dS})} \right)^2, \qquad {\cal N}_{\rm AdS}^{\rm (sc)} = \left( \frac{\sinh\pi \bar{\mu}(R_{\rm AdS})}{\cosh\pi \kappa(R_{\rm AdS})} \right)^2,
\end{eqnarray}
where
\begin{eqnarray}
\bar{\mu}(R) = \sqrt{\mu^2(R) - \Bigl(\frac{1}{2} \Bigr)^2} \,.
\end{eqnarray}
In fact, the reciprocal relation holds
\begin{eqnarray}
{\cal N}_{\rm dS}^{\rm (sc)} (R) {\cal N}_{\rm AdS}^{\rm (sc)} (R) = 1,
\label{sc_rec}
\end{eqnarray}
with $R = 2 H^2$ or $R = -2 H^2$. Note that the mean numbers between scalars and spinors are connected by the uniform electric field with replacing $\bar{\mu}$ with $\mu + i/2$, i.e. $|\sinh\pi\bar{\mu}| = \cosh\pi\mu$. In a strong electric field ($\mu \gg 1$), i.e. $\bar\mu \simeq \mu - 1/8 \mu$, scalar production is larger than spinor production in the dS space while spinor production is larger than scalar production in the AdS space, which are approximately given as, ${\cal N}_{\rm dS}^{\rm (sc)} \simeq {\cal N}_{\rm dS}^{\rm (sp)} \times \mathrm{e}^{\pi /8 \mu}$ and ${\cal N}_{\rm AdS}^{\rm (sc)} \simeq {\cal N}_{\rm AdS}^{\rm (sp)} \times \mathrm{e}^{-\pi/8 \mu}$. But there is no bosonic amplification and the mean numbers can never exceed one.

In the planar ${\rm (A)dS}_2$, the mean numbers of scalars with the sign of $\bar{k}$ ($\bar{\omega}$) included in~\cite{Cai:2014qba},
\begin{eqnarray}
{\cal N}_{\rm dS}^{\rm (sc)} = \frac{\cosh(\pi \bar{\mu} + \mathrm{sgn}(\bar{k}) \, \pi \kappa ) }{\mathrm{e}^{\pi \bar{\mu} - \mathrm{sgn}(\bar{k}) \, \pi \kappa} \sinh(2 \pi \bar{\mu})}, \qquad {\cal N}_{\rm AdS}^{\rm (sc)} = \frac{\mathrm{e}^{\pi \bar{\mu} - \mathrm{sgn}(\bar{\omega}) \, \pi \kappa} \sinh(2 \pi \bar{\mu})}{\cosh(\pi \bar{\mu} + \mathrm{sgn}(\bar{\omega}) \, \pi \kappa)},
\end{eqnarray}
satisfy the reciprocal relation~(\ref{sc_rec}). As in the global coordinates, the scalar production is larger (smaller) than the spinor production in the dS (AdS) space.

\section{Phase-Integral Formulation of Pair Production}\label{phase_integral}
We use the phase-integral formulation~\cite{Chen:2012zn, Kim:2013iua, Kim:2013cka, Chen:2023swn} to explain the leading behavior of Schwinger pair production of spinor and scalar. To do so, we consider the second-order formulation in a canonical form for spinor and scalar field in the global $\mathrm{dS}_2$:
\begin{equation} \label{can_eq_dS}
\left[ \partial_{\tau}^2 + \omega_{k \sigma}^2(\tau) \right] \tilde{\phi}_{k \sigma}(\tau) = 0,
\end{equation}
where $\psi_{k \sigma} = \tilde{\phi}_{k \sigma} (\tau)/\sqrt{\cosh \tau}$ and the frequency square is
\begin{equation}
\omega_{k \sigma}^2(\tau) = \mu^2 + 2 k \bar\kappa \tanh\tau \, \mathrm{sech}\,\tau - \left( \bar\kappa^2 - k^2 + 1/4 \right) \mathrm{sech}^2\,\tau,
\label{mode eq}
\end{equation}
in which $\bar{\kappa} = \kappa + i \sigma/2$ and $\sigma = 1, -1$ for spinor $\uparrow, \downarrow$, $\sigma = 0$ for scalar, and $k$ is a half-integer for spinor and an integer for scalar. The phase of the positive frequency solution of~(\ref{mode eq}) is given by $\tilde{\phi}_{k \sigma}(\tau) = \exp\left( -i \int d\tau \omega_{k \sigma}(\tau) \right)$. In the phase-integral formulation the leading mean number~\cite{Kim:2013cka} is given by
\begin{equation}
{\cal N} = \sum_{C_J} \exp\Bigl[ - 2 \pi \sum_i \mathrm{Res}(\omega_{k \sigma}(\tau_{J}^i)) \Bigr],
\end{equation}
where $C_J$ denotes all possible contours of winding number one that start from an initial point, and $\tau_{J}^i$ are poles enclosed by $C_J$.
Using the conformal mapping $z = \sinh(\tau)$, the contour integral becomes
\begin{eqnarray}
\oint \frac{dz}{z^2 + 1} \sqrt{\mu^2 (z^2 + 1) + 2 k \bar{\kappa} z - \Bigl( \bar{\kappa}^2 - k^2 + 1/4 \Bigr)} = \oint dz \frac{\mu}{z^2 + 1} \sqrt{(z - z_1) (z - z_2)},
\end{eqnarray}
where $z_1, z_2$ are branch points of the contour integration. There are three simple poles at $z = \pm i$ and $z = \infty$ and the associated residues are $\mathrm{Res}(\omega_{k\sigma}(\infty)) = \mu$ and $\mathrm{Res}(\omega_{k\sigma}(\pm i)) = \pm \sqrt{(\bar\kappa \mp i k)^2 - 1/4}/2 \approx \pm \kappa/2$ for large $\kappa$. We can make $\omega_{k \sigma}(z)$ an analytical function by introducing branch-cuts~\cite{Markushevich} and the contours are chosen by the causality. The simple pole at infinity corresponding to the physical in- and out-modes should be always included. Moreover, for a closed contour in $\tau$-coordinates, for example, from $\tau = -\infty$ to $\tau = \infty$ and back, actually corresponds to a ``double'' contour in $z = \sinh(\tau)$ coordinates. Therefore, the contribution of the poles at $z = \pm i$ should be doubly counted with their residues. The contours for leading contributions give
\begin{equation}
{\cal N}_{k \sigma} \simeq \mathrm{e}^{- 2 \pi \mu} \left( 1 + \mathrm{e}^{2 \pi \kappa} + \mathrm{e}^{- 2 \pi \kappa} + \mathrm{e}^{2 \pi (\kappa - \kappa)} \right) = \mathrm{e}^{- 2 \pi (\mu - \kappa)} + 2 \mathrm{e}^{- 2 \pi \mu} + \mathrm{e}^{- 2 \pi (\mu + \kappa)},
\end{equation}
where the first term comes from a contour excluding both $\pm i$ and the second and third terms come from contours enclosing only one pole $i$ or $- i$, and the last term from the contour enclosing both poles $\pm i$. The result is consistent with the common leading term for scalars~(\ref{scalar_ds}) and spinor~(\ref{spinor_ds}).

Similarly, the second-order formulation in a canonical form for spinor in the planar $\mathrm{dS}_2$, with $\psi_{k\sigma} = \mathrm{e}^{-\tau/2} \tilde\phi_{k\sigma}(\tau)$, has the frequency square
\begin{eqnarray}
\omega_{k \sigma}^2(\tau) = \mu^2 + 2  \bar{k} \bar{\kappa} \mathrm{e}^{-\tau} + \bar{k}^2 \mathrm{e}^{-2 \tau}.
\end{eqnarray}
With the conformal mapping $z = \exp(-\tau)$, the contour integral of the phase
\begin{eqnarray}
\oint dz \frac{\sqrt{\bar{k}^2 z^2 + 2 \bar{k} \bar{\kappa} z + \mu^2}}{z},
\end{eqnarray}
which has two branch points $z_{\pm} = (- \bar{\kappa} \pm \sqrt{\bar{\kappa}^2 - \mu^2 })/\bar{k}$ and two simple poles at $z = 0$ and $z = \infty$ with residues $\mathrm{Res}(\omega_{k\sigma}(0)) = \mu$ and $\mathrm{Res}(\omega_{k\sigma}(\infty)) = - \mathrm{sgn}(\bar{k}) \, \kappa$, respectively. The contribution from the contour enclosing both poles leads to
\begin{eqnarray}
{\cal N}_{\mathrm{dS}} \simeq \mathrm{e}^{- 2 \pi (\mu - \mathrm{sgn} (\bar{k}) \kappa)}.
\end{eqnarray}

In the global coordinates of AdS$_2$, the second-order formulation for each component $\psi_{\omega \sigma} = \tilde{\phi}_{\omega \sigma}(\zeta)/\sqrt{\cosh\zeta}$ takes the canonical form
\begin{eqnarray}  \label{can_eq_AdS}
\left[ \partial_{\zeta}^2 + k_{\omega \sigma}^2(\zeta) \right] \tilde{\phi}_{\omega \sigma}(\zeta) = 0,
\end{eqnarray}
where the momentum square is
\begin{equation}
k_{\omega \sigma}^2(\zeta) = \mu^2 + 2 \omega \bar{\kappa}^* \tanh \zeta \, \mathrm{sech}\,\zeta - \left( \bar{\kappa}^{*2} - \omega^2 + 1/4 \right) \mathrm{sech}^2\,\zeta.
\end{equation}
In the planar coordinates the momentum square takes the form
\begin{eqnarray}
k_{\omega \sigma}^2(\zeta) = \mu^2 + 2 \bar{\omega} \bar{\kappa}^* \mathrm{e}^{-\zeta} + \bar{\omega}^2 \mathrm{e}^{-2 \zeta}.
\end{eqnarray}
The phase $\tilde{\phi}_{\omega \sigma}(\zeta) = \exp\left( i \int d \zeta k_{\omega \sigma} (\zeta) \right)$ decays due to residues from simple poles at $z = \sinh(\zeta) = \pm i$ and another pole $z = \infty$ in the global coordinates and simple poles at $z = \exp(- \zeta) = 0$ and $z = \infty$ in the planar coordinates. These poles contribute to the Boltzmann factor ${\cal N}_{\rm AdS} \simeq \mathrm{e}^{- 2 \pi (\kappa - \mu)}$ of the mean number of spinors or scalars in the global coordinates and ${\cal N}_{\rm AdS} \simeq \mathrm{e}^{2 \pi (\mu - \mathrm{sgn}(\bar{\omega}) \kappa)}$ in the planar coordinates.

\section{Conclusion} \label{conclusion}
We have exactly obtained the Schwinger production of spinors (spin-1/2 fermions) and scalars (spin-0 bosons) in a uniform electric field in the global coordinates of (A)dS$_2$ and compared the results with those in the planar coordinates. The Dirac and Klein-Gordon equations in a uniform electric field respect the symmetry SO(2,1) (SO(1,2)) in (A)dS$_2$ and have solutions in terms of hypergeometric functions, from which the Bogoliubov relations are found between the in- and out-vacua with asymptotic WKB (adiabatic) in- and out-states, and the mean numbers of Schwinger pair production of spinors and scalars are calculated in the in-out formalism. In the global dS$_2$ space, for instance, the mean number~(\ref{spinor_ds}) gives rise to the spinor production rate per unit two volume, which may have an invariant expression in terms of the Maxwell scalar, ${\cal F} = E^2/2$, and the scalar curvature, $R = 2 H^2$,
\begin{eqnarray}
{\cal R}^{\rm (sp)}_{\rm dS} = {\cal D} \Biggl(\frac{\cosh 2 \pi q \frac{\sqrt{2{\cal F}}}{R}}{\cosh 2 \pi q \sqrt{\frac{2 {\cal F}}{R^2} + \frac{m^2}{2 q^2 R}}} \Biggr)^2,
\label{sp_rate}
\end{eqnarray}
where ${\cal D} = g (H^2/2 \pi) \mu = g (q \sqrt{2 {\cal F}}/2 \pi) \times \sqrt{1 + m^2 R/4 q^2 {\cal F}}$ is the density of states with the spin multiplicity ($g = 2$) in an electric field that is uniformly distributed in the global geometry~\cite{Kim:2016nyz}. The formula recovers the limits both of Minkowski space with the electric field and pure dS$_2$ space without any electric field.

First, in the Minkowski limit ($R = 2H^2 = 0$), the spinor production rate recovers the Schwinger formula in two dimensions~\cite{Cohen:2008wz}
\begin{eqnarray}
{\cal R}^{\rm (sp)}_{\rm Min} = \frac{g qE}{2 \pi} \mathrm{e}^{- \pi \frac{m^2}{q E}}.
\end{eqnarray}
Second, in the limit of zero electric field or charge neutral particle, the spinor production rate recovers the pure dS$_2$ in global coordinates~\cite{Jiang:2020evx} except for the density of states ${\cal D} = g mH/2 \pi$, which is required for the effective action in pure dS$_2$. However, there is no production of spinors in the AdS$_2$ without electric field, and furthermore, the BF bound $qE/H^2 \leq m/H$ prohibits weak electric fields from producing any spinor.

The main results of this paper show the combined effect of an electric field and spacetime curvature for pair production in a nontrivial way. The crucial factor is the ratio of the Maxwell scalar to the scalar curvature square or the ratio of the dS radius to the electric length ($l_{\rm E} = 1/\sqrt{qE}$), as shown in~(\ref{sp_rate}). Another factor is the ratio of the dS radius to the local Compton wavelength of charge ($1/m$). Except for the two limits of the electric field or the curvature dominance, the QED and curvature effects are strongly coupled and intertwined in producing pairs. An intuitive way to understand the emission formula is to use the relevant temperatures: the Unruh temperature for charge acceleration by the electric field and the Gibbons-Hawking temperature in dS space. Provided that the condition $l_{\rm E} < l_{\rm dS}$ for pair production is satisfied, the mean number~(\ref{dS-spinor}) has the leading Boltzmann factor
\begin{eqnarray}
{\cal N}^{\rm (sp)}_{\rm dS} = \mathrm{e}^{- (m/T_{\rm GH}) \times n_{\rm S}}, \quad n_{\rm S} = \frac{1}{\sqrt{1 + (T_{\rm U}/T_{\rm GH})^2} + T_{\rm U}/ T_{\rm GH}},
\end{eqnarray}
in which the Gibbons-Hawking radiation is enhanced by an exponential folding factor:
\begin{eqnarray}
n_{\rm S} =
\begin{cases}
1- \frac{T_{\rm U}}{T_{\rm GH}} + \frac{1}{2} \Bigl( \frac{T_{\rm U}}{T_{\rm GH}} \Bigr)^2 - \frac{1}{8} \Bigl( \frac{T_{\rm U}}{T_{\rm GH}} \Bigr)^4 + {\cal O} \Bigl( \frac{T_{\rm U}}{T_{\rm GH}} \Bigr)^6, \quad & (T_{\rm GH} > T_{\rm U}),
\\
\frac{1}{2} \frac{T_{\rm GH}}{T_{\rm U}} - \frac{1}{8} \Bigl( \frac{T_{\rm GH}}{T_{\rm U}} \Bigr)^3 + \frac{1}{16} \Bigl( \frac{T_{\rm GH}}{T_{\rm U}} \Bigr)^5 + {\cal O} \Bigl( \frac{T_{\rm U}}{T_{\rm GH}} \Bigr)^7, \quad &(T_{\rm U} > T_{\rm GH}).
\end{cases}
\end{eqnarray}
When $T_{\rm GH} \gg T_{\rm U}$, the Boltmann factor may be further interpreted as Gibbons-Hawking radiation of massive charge with a chemical potential $qE l_{\rm dS}$. In the opposite case of $T_{\rm U} \gg T_{\rm GH}$ when the Schwinger effect dominates over the Gibbons-Hawking radiation, the Boltzmann factor may be written as ${\cal N}^{\rm (sp)}_{\rm dS} = \mathrm{e}^{- (m/2 T_{\rm U}) \times n_{\rm GH}}$, in which the e-folding factor $n_{\rm GH} = 2/(\sqrt{1 + (T_{\rm GH}/ T_{\rm U})^2} + 1)$ gives the curvature effect on Schwinger pair production. This is a novel feature of strong coupling of gauge field and spacetime curvature, which may affect the Schwinger pair production in the early universe.

We have observed that the exact mean numbers for Schwinger pair production of spinors and scalars in a uniform electric field satisfy the reciprocal relation between dS$_2$ and AdS$_2$ in the global coordinates and planar coordinates, respectively, when the scalar curvature is analytically continued. This means that the mean number in AdS$_2$ can be obtained from that of dS$_2$ by analytically continuing the scalar curvature from $R = 2 H^2$ to $R = -2 H^2$ and vice versa, which holds for spinors and scalars and the global coordinates and planar coordinates, respectively. It is originated from scattering and tunneling boundary conditions on a quantum field. A physical implication of the reciprocal relation is that QED may connect dS$_2$ with AdS$_2$ space at the one-loop level. The QED effective action will be studied in the future.

\acknowledgments
C.~M.~C. and S.~P.~K. thank the participants of the joint program [APCTP-2025-J01] held at APCTP, Pohang, Korea for fruitful discussions. The work of C.~M.~C. was supported by the National Science Council of the R.O.C. (Taiwan) under the grant NSTC 113-2112-M-008-027. The work of S.~P.~K. was supported in part by the Institute of Basic Science (Grant No. IBSR038-D1).

\begin{appendix}

\section{Bogoliubov Coefficients of Spinor in Global dS$_2$ } \label{App_dS}
We show the detailed calculations that lead to the Bogoliubov coefficients. For the spinor solutions~\eqref{sol_spinor_GdS1} and~\eqref{sol_spinor_GdS2}, in order to classify the positive and negative frequency modes for the in- and out-states, one needs the asymptotic behaviours of $\phi_{\uparrow1}$ and $\phi_{\uparrow2}$ for $\tau \to - \infty$
\begin{eqnarray}
\phi_{\uparrow1} &\to& A_1 \left( 1/2 \right)^{-1/2 - i \kappa + 2 i \mu} \mathrm{e}^{(-i k - \kappa + \mu) \pi/2} \mathrm{e}^{\tau/2 - i \mu \tau} + B_1 \left( 1/2 \right)^{-1/2 - i \kappa - 2 i \mu} \mathrm{e}^{(-i k - \kappa - \mu) \pi/2} \mathrm{e}^{\tau/2 + i \mu \tau},
\nonumber\\
\phi_{\uparrow2} &\to& A_2 \left( 1/2 \right)^{-3/2 + i \kappa + 2 i \mu} \mathrm{e}^{(i + i k + \kappa + \mu) \pi/2} \mathrm{e}^{\tau/2 - i \mu \tau} + B_2 \left( 1/2 \right)^{-3/2 + i \kappa - 2 i \mu} \mathrm{e}^{(i + i k + \kappa - \mu) \pi/2} \mathrm{e}^{\tau/2 + i \mu \tau},
\end{eqnarray}
and similarly for $\tau \to \infty$
\begin{eqnarray}
\phi_{\uparrow1} &\to& A_1 \left( 1/2 \right)^{-1/2 - i \kappa + 2 i \mu} \mathrm{e}^{(i k + \kappa - \mu) \pi/2} \mathrm{e}^{-\tau/2 + i \mu \tau} + B_1 \left( 1/2 \right)^{-1/2 - i \kappa - 2 i \mu} \mathrm{e}^{(i k + \kappa + \mu) \pi/2} \mathrm{e}^{-\tau/2 - i \mu \tau},
\nonumber\\
\phi_{\uparrow2} &\to& A_2 \left( 1/2 \right)^{-3/2 + i \kappa + 2 i \mu} \mathrm{e}^{(-i - i k - \kappa - \mu) \pi/2} \mathrm{e}^{-\tau/2 + i \mu \tau} + B_2 \left( 1/2 \right)^{-3/2 + i \kappa - 2 i \mu} \mathrm{e}^{(-i - i k - \kappa + \mu) \pi/2} \mathrm{e}^{-\tau/2 - i \mu \tau},
\end{eqnarray}
where the four coefficients $A_1, A_2, B_1$ and $B_2$ are given by
\begin{eqnarray}
&& A_1 = \frac{\Gamma(1/2 - k + i \kappa) \Gamma(i 2 \mu)}{\Gamma(i \kappa + i \mu) \Gamma(1/2 - k + i \mu)}, \qquad B_1 = \frac{\Gamma(1/2 - k + i \kappa) \Gamma(- i 2 \mu)}{\Gamma(i \kappa - i \mu) \Gamma(1/2 - k - i \mu)},
\nonumber\\
&& A_2 = \frac{\Gamma(3/2 + k - i \kappa) \Gamma(i 2 \mu)}{\Gamma(1 - i \kappa + i \mu) \Gamma(1/2 + k + i \mu)}, \qquad B_2 = \frac{\Gamma(3/2 + k - i \kappa) \Gamma(- i 2 \mu)}{\Gamma(1 - i \kappa - i \mu) \Gamma(1/2 + k - i \mu)}.
\end{eqnarray}
Then, these coefficients have the following relations, which are useful for calculating the Bogoliubov coefficients,
\begin{equation}
\frac{A_2}{A_1} \frac{\tilde A_2'}{\tilde A_1'} = \frac{B_2}{B_1} \frac{\tilde B_2'}{\tilde B_1'} = - \frac14 \frac{c_2}{c_1},
\end{equation}
where $A' = A(k \to -k)$ and $\tilde A = A(\kappa \to -\kappa)$.
By using the relations, with half-integer $k$, one can straightforwardly confirm the following relations
\begin{eqnarray}
&& A_1 B_2 - A_2 B_1 = \frac{i (1 + 2 k - 2 i \kappa)}{4 \mu}, \qquad A_1 B_2 \mathrm{e}^{\pi \kappa - \pi \mu} - A_2 B_1 \mathrm{e}^{- \pi \kappa - \pi \mu} = \frac{i (1 + 2 k - 2 i \kappa)}{4 \mu} \frac{\cosh\pi \kappa}{\cosh\pi \mu},
\nonumber\\
&& B_1 B_2 = i \mathrm{e}^{-i k \pi} \frac{\pi (1 + 2 k - 2 i \kappa)}{2 (\mu + \kappa) \sinh\pi \kappa} \frac{\Gamma(-2 i \mu)^2}{\Gamma(-i \kappa - i \mu) \Gamma(i \kappa - i \mu) \Gamma(1/2 + k - i \mu) \Gamma(1/2 - k - i \mu)}.
\end{eqnarray}

\section{Bogoliubov Coefficients of Spinor in Global AdS$_2$ } \label{App_AdS}
As in the case of dS$_2$, we can find the Bogoliubov coefficients, but the difference is the tunneling boundary condition from AdS$_2$ space. The asymptotic behaviours of $\phi_{\uparrow1}$ and $\phi_{\uparrow2}$~\eqref{sol_spinor_GAdS1} are: for $\zeta \to - \infty$
\begin{eqnarray}
\phi_{\uparrow1} &\to& A_1 \left( 1/2 \right)^{-1/2 + i \kappa + 2 i \mu} \mathrm{e}^{(i \omega + \kappa + \mu) \pi/2} \mathrm{e}^{\zeta/2 - i \mu \zeta} + B_1 \left( 1/2 \right)^{-1/2 + i \kappa - 2 i \mu} \mathrm{e}^{(i \omega + \kappa - \mu) \pi/2} \mathrm{e}^{\zeta/2 + i \mu \zeta},
\nonumber\\
\phi_{\uparrow2} &\to& A_2 \left( 1/2 \right)^{-3/2 - i \kappa + 2 i \mu} \mathrm{e}^{(i - i \omega - \kappa + \mu) \pi/2} \mathrm{e}^{\zeta/2 - i \mu \zeta} + B_2 \left( 1/2 \right)^{-3/2 - i \kappa - 2 i \mu} \mathrm{e}^{(i - i \omega - \kappa - \mu) \pi/2} \mathrm{e}^{\zeta/2 + i \mu \zeta},
\end{eqnarray}
and for $\zeta \to \infty$
\begin{eqnarray}
\phi_{\uparrow1} &\to& A_1 \left( 1/2 \right)^{-1/2 + i \kappa + 2 i \mu} \mathrm{e}^{(-i \omega - \kappa - \mu) \pi/2} \mathrm{e}^{-\zeta/2 + i \mu \zeta} + B_1 \left( 1/2 \right)^{-1/2 + i \kappa - 2 i \mu} \mathrm{e}^{(-i \omega - \kappa + \mu) \pi/2} \mathrm{e}^{-\zeta/2 - i \mu \zeta},
\nonumber\\
\phi_{\uparrow2} &\to& A_2 \left( 1/2 \right)^{-3/2 - i \kappa + 2 i \mu} \mathrm{e}^{(-i + i \omega + \kappa - \mu) \pi/2} \mathrm{e}^{-\zeta/2 + i \mu \zeta} + B_2 \left( 1/2 \right)^{-3/2 - i \kappa - 2 i \mu} \mathrm{e}^{(-i + i \omega + \kappa + \mu) \pi/2} \mathrm{e}^{-\zeta/2 - i \mu \zeta},
\end{eqnarray}
where the four coefficients are
\begin{eqnarray}
&& A_1 = \frac{\Gamma(1/2 + \omega - i \kappa) \Gamma(i 2 \mu)}{\Gamma(-i \kappa + i \mu) \Gamma(1/2 + \omega + i \mu)}, \qquad B_1 = \frac{\Gamma(1/2 + \omega - i \kappa) \Gamma(- i 2 \mu)}{\Gamma(-i \kappa - i \mu) \Gamma(1/2 + \omega - i \mu)},
\nonumber\\
&& A_2 = \frac{\Gamma(3/2 - \omega + i \kappa) \Gamma(i 2 \mu)}{\Gamma(1 + i \kappa + i \mu) \Gamma(1/2 - \omega + i \mu)}, \qquad B_2 = \frac{\Gamma(3/2 - \omega + i \kappa) \Gamma(- i 2 \mu)}{\Gamma(1 + i \kappa - i \mu) \Gamma(1/2 - \omega - i \mu)}.
\end{eqnarray}
The following relations hold among these four coefficients, which will be used in finding the Bogoliubov coefficients,
\begin{equation}
\frac{A_2}{A_1} \frac{\tilde A_2'}{\tilde A_1'} = \frac{B_2}{B_1} \frac{\tilde B_2'}{\tilde B_1'} = - \frac14 \frac{c_2}{c_1},
\end{equation}
where $A' = A(\omega \to -\omega)$ and $\tilde A = A(\kappa \to -\kappa)$.
By using the following relations, with half-integer $\omega$, we obtain
\begin{eqnarray}
&& A_1 B_2 - A_2 B_1 = \frac{i (1 - 2 \omega + 2 i \kappa)}{4 \mu}, \qquad A_1 B_2 \mathrm{e}^{\pi \kappa + \pi \mu} - A_2 B_1 \mathrm{e}^{- \pi \kappa + \pi \mu} = \frac{i (1 - 2 \omega + 2 i \kappa)}{4 \mu} \frac{\cosh\pi \kappa}{\cosh\pi \mu},
\nonumber\\
&& B_1 B_2 = i \mathrm{e}^{i \omega \pi} \frac{\pi (1 - 2 \omega + 2 i \kappa)}{2 (\kappa - \mu) \sinh\pi \kappa} \frac{\Gamma(-2 i \mu)^2}{\Gamma(-i \kappa - i \mu) \Gamma(i \kappa - i \mu) \Gamma(1/2 + \omega - i \mu) \Gamma(1/2 - \omega - i \mu)}.
\end{eqnarray}

\end{appendix}


\begin{thebibliography}{99}

\bibitem{Griffiths:2009dfa}
J.~B.~Griffiths and J.~Podolsky,
``Exact Space-Times in Einstein's General Relativity,''
(Cambridge University Press, 2009)

\bibitem{Chernikov:1968zm}
N.~A.~Chernikov and E.~A.~Tagirov,
``Quantum theory of scalar fields in de Sitter space-time,''
Ann. Inst. H. Poincare A Phys. Theor. \textbf{9} (1968), 109

\bibitem{Bousso:2001mw}
R.~Bousso, A.~Maloney and A.~Strominger,
``Conformal vacua and entropy in de Sitter space,''
Phys. Rev. D \textbf{65} (2002), 104039
[arXiv:hep-th/0112218 [hep-th]].

\bibitem{Avis:1977yn}
S.~J.~Avis, C.~J.~Isham and D.~Storey,
``Quantum Field Theory in anti-De Sitter Space-Time,''
Phys. Rev. D \textbf{18} (1978), 3565

\bibitem{Spradlin:2001pw}
M.~Spradlin, A.~Strominger and A.~Volovich,
``Les Houches lectures on de Sitter space,''
[arXiv:hep-th/0110007 [hep-th]].

\bibitem{Maldacena:2016upp}
J.~Maldacena, D.~Stanford and Z.~Yang,
``Conformal symmetry and its breaking in two dimensional Nearly Anti-de-Sitter space,''
PTEP \textbf{2016} (2016), 12C104
[arXiv:1606.01857 [hep-th]].

\bibitem{Joung:2006gj}
E.~Joung, J.~Mourad and R.~Parentani,
``Group theoretical approach to quantum fields in de Sitter space. I. The Principle series,''
JHEP \textbf{08} (2006), 082
[arXiv:hep-th/0606119 [hep-th]].



\bibitem{Witten:1998qj}
E.~Witten,
``Anti de Sitter space and holography,''
Adv. Theor. Math. Phys. \textbf{2} (1998), 253
[arXiv:hep-th/9802150 [hep-th]].

\bibitem{Anninos:2019oka}
D.~Anninos, D.~M.~Hofman and J.~Kruthoff,
``Charged Quantum Fields in AdS$_2$,''
SciPost Phys. \textbf{7} (2019), 054
[arXiv:1906.00924 [hep-th]].

\bibitem{Bardeen:1999px}
J.~M.~Bardeen and G.~T.~Horowitz,
``The Extreme Kerr throat geometry: A Vacuum analog of $\mathrm{AdS}_2 \times \mathrm{S}^2$,''
Phys. Rev. D \textbf{60} (1999), 104030
[arXiv:hep-th/9905099 [hep-th]].

\bibitem{Ortaggio:2002bp}
M.~Ortaggio and J.~Podolsky,
``Impulsive waves in electrovac direct product space-times with Lambda,''
Class. Quant. Grav. \textbf{19} (2002), 5221-5227
[arXiv:gr-qc/0209068 [gr-qc]].

\bibitem{Kunduri:2013gce}
H.~K.~Kunduri and J.~Lucietti,
``Classification of near-horizon geometries of extremal black holes,''
Living Rev. Rel. \textbf{16} (2013), 8
[arXiv:1306.2517 [hep-th]].

\bibitem{Hawking:1975vcx}
S.~W.~Hawking,
``Particle Creation by Black Holes,''
Commun. Math. Phys. \textbf{43} (1975), 199-220
[erratum: Commun. Math. Phys. \textbf{46} (1976), 206]

\bibitem{Schwinger:1951nm}
J.~S.~Schwinger,
``On gauge invariance and vacuum polarization,''
Phys. Rev. \textbf{82} (1951), 664-679

\bibitem{Parker:1968mv}
L.~Parker,
``Particle creation in expanding universes,''
Phys. Rev. Lett. \textbf{21} (1968), 562-564

\bibitem{Gibbons:1977mu}
G.~W.~Gibbons and S.~W.~Hawking,
``Cosmological Event Horizons, Thermodynamics, and Particle Creation,''
Phys. Rev. D \textbf{15} (1977), 2738-2751

\bibitem{Gibbons:1975}
G.~W.~Gibbons,
``Vacuum polarization and the spontaneous loss of charge by black holes,''
Commun. Math. Phys. \textbf{44} (1975), 245-264

\bibitem{Page:1977um}
D.~N.~Page,
``Particle Emission Rates from a Black Hole. 3. Charged Leptons from a Nonrotating Hole,''
Phys. Rev. D \textbf{16} (1977), 2402-2411


\bibitem{Hiscock:1990ex}
W.~A.~Hiscock and L.~D.~Weems,
``Evolution of Charged Evaporating Black Holes,''
Phys. Rev. D \textbf{41} (1990), 1142


\bibitem{Montero:2019ekk}
M.~Montero, T.~Van Riet and V.~Venken,
``Festina Lente: EFT Constraints from Charged Black Hole Evaporation in de Sitter,''
JHEP \textbf{01} (2020), 039
[arXiv:1910.01648 [hep-th]].

\bibitem{Brown:2024ajk}
A.~R.~Brown, L.~V.~Iliesiu, G.~Penington and M.~Usatyuk,
``The evaporation of charged black holes,''
[arXiv:2411.03447 [hep-th]].


\bibitem{Chen:2012zn}
C.-M.~Chen, S.~P.~Kim, I.-C.~Lin, J.-R.~Sun and M.-F.~Wu,
``Spontaneous Pair Production in Reissner-Nordstrom Black Holes,''
Phys. Rev. D \textbf{85} (2012), 124041
[arXiv:1202.3224 [hep-th]].

\bibitem{Chen:2020mqs}
C.-M.~Chen and S.~P.~Kim,
``Schwinger Effect from Near-extremal Black Holes in (A)dS Space,''
Phys. Rev. D \textbf{101} (2020), 085014
[arXiv:2002.00394 [hep-th]].

\bibitem{Chen:2023swn}
C.-M.~Chen, C.-C.~Huang, S.~P.~Kim and C.-Y.~Wei,
``Catastrophic emission of charges from near-extremal Nariai black holes,''
Phys. Rev. D \textbf{110} (2024), 085020
[arXiv:2309.00218 [hep-th]].

\bibitem{Chen:2021jwy}
C.-M.~Chen and S.~P.~Kim,
``Dyon production from near-extremal Kerr-Newman-(anti-)de Sitter black holes,''
Eur. Phys. J. C \textbf{83} (2023), 219
[arXiv:2111.14124 [hep-th]].

\bibitem{Chen:2024ctf}
C.-M.~Chen, C.-C.~Huang, S.~P.~Kim and C.-Y.~Wei,
``Catastrophic emission of charges from near-extremal charged Nariai black holes. II. Rotation effect,''
Phys. Rev. D \textbf{111} (2025), 065017
[arXiv:2408.12343 [hep-th]].

\bibitem{Lin:2024jug}
P.~Lin and G.~Shiu,
``Schwinger effect of extremal Reissner-Nordstr\"om black holes,''
JHEP \textbf{06} (2025), 017
[arXiv:2409.02197 [hep-th]].

\bibitem{Garriga:1994bm}
J.~Garriga,
``Pair production by an electric field in (1+1)-dimensional de Sitter space,''
Phys. Rev. D \textbf{49} (1994), 6343-6346

\bibitem{Kim:2008xv}
S.~P.~Kim and D.~N.~Page,
``Schwinger Pair Production in dS(2) and AdS(2),''
Phys. Rev. D \textbf{78} (2008), 103517
[arXiv:0803.2555 [hep-th]].


\bibitem{Pioline:2005pf}
B.~Pioline and J.~Troost,
``Schwinger pair production in AdS(2),''
JHEP \textbf{03} (2005), 043
[arXiv:hep-th/0501169 [hep-th]].

\bibitem{Mottola:1984ar}
E.~Mottola,
``Particle Creation in de Sitter Space,''
Phys. Rev. D \textbf{31} (1985), 754-766

\bibitem{Cooper:1994eh}
F.~Cooper, A.~Khare and U.~Sukhatme,
``Supersymmetry and quantum mechanics,''
Phys. Rept. \textbf{251} (1995), 267-385
[arXiv:hep-th/9405029 [hep-th]].

\bibitem{Kim:2013iua}
S.~P.~Kim,
``New Geometric Transition as Origin of Particle Production in Time-Dependent Backgrounds,''
Phys. Lett. B \textbf{725} (2013), 500-503
[arXiv:1306.5549 [hep-th]].

\bibitem{Kim:2013cka}
S.~P.~Kim,
``Geometric Origin of Stokes Phenomenon for de Sitter Radiation,''
Phys. Rev. D \textbf{88} (2013), 044027
[arXiv:1307.0590 [hep-th]].

\bibitem{Jiang:2020evx}
J.~Jiang,
``Scalar and Spinor Effective Actions in Global de Sitter,''
JHEP \textbf{06} (2020), 037
[arXiv:2004.06251 [hep-th]].

\bibitem{Schaub:2023scu}
V.~Schaub,
``Spinors in (Anti-)de Sitter Space,''
JHEP \textbf{09} (2023), 142
[arXiv:2302.08535 [hep-th]].

\bibitem{Nikishov:1970br}
A.~I.~Nikishov,
``Barrier scattering in field theory removal of klein paradox,''
Nucl. Phys. B \textbf{21} (1970), 346-358

\bibitem{Kim:2008yt}
S.~P.~Kim, H.~K.~Lee and Y.~Yoon,
``Effective Action of Scalar QED in Electric Field Backgrounds,''
Phys. Rev. D \textbf{78} (2008), 105013
[arXiv:0807.2696 [hep-th]].

\bibitem{Kim:2009pg}
S.~P.~Kim, H.~K.~Lee and Y.~Yoon,
``Effective Action of QED in Electric Field Backgrounds II. Spatially Localized Fields,''
Phys. Rev. D \textbf{82} (2010), 025015
[arXiv:0910.3363 [hep-th]].

\bibitem{Cai:2014qba}
R.-G.~Cai and S.~P.~Kim,
``One-Loop Effective Action and Schwinger Effect in (Anti-) de Sitter Space,''
JHEP \textbf{09} (2014), 072
[arXiv:1407.4569 [hep-th]].


\bibitem{Villalba:1995za}
V.~M.~Villalba,
``Creation of spin 1/2 particles by an electric field in de Sitter space,''
Phys. Rev. D \textbf{52} (1995), 3742-3745
[arXiv:hep-th/9507021 [hep-th]].

\bibitem{Haouat:2012ik}
S.~Haouat and R.~Chekireb,
``Comment on \textquotedblleft{}Creation of spin 1/2 particles by an electric field in de Sitter space\textquotedblright{},''
Phys. Rev. D \textbf{87} (2013), 088501
[arXiv:1207.4342 [hep-th]].

\bibitem{Stahl:2015cra}
C.~Stahl and S.~Eckhard,
``Semiclassical fermion pair creation in de Sitter spacetime,''
AIP Conf. Proc. \textbf{1693} (2015), 050005
[arXiv:1507.01401 [hep-th]].

\bibitem{Stahl:2015gaa}
C.~Stahl, E.~Strobel and S.~S.~Xue,
``Fermionic current and Schwinger effect in de Sitter spacetime,''
Phys. Rev. D \textbf{93} (2016), 025004
[arXiv:1507.01686 [gr-qc]].

\bibitem{Botshekananfard:2019zak}
M.~Botshekananfard and E.~Bavarsad,
``Induced energy-momentum tensor of a Dirac field in 2D de Sitter QED,''
Phys. Rev. D \textbf{101} (2020), 085011
[arXiv:1911.10588 [hep-th]].

\bibitem{Kim:2022nsx}
S.~P.~Kim, W.~Y.~P.~Hwang and T.~C.~Wang,
``Schwinger mechanism in global $dS_2$ and $AdS_2$ space,''
Chin. J. Phys. \textbf{77} (2022), 2073-2077


\bibitem{1945RvMP...17..343C}
H.~B. Casimir,
``On Onsager's Principle of Microscopic Reversibility,''
Rev. Mod. Phys. \textbf{17} (1945), 343-350

\bibitem{1969PhRv..183.1057P}
L.~Parker,
``Quantized Fields and Particle Creation in Expanding Universes. I''
Phys. Rev. \textbf{183} (1969), 1057–1068

\bibitem{1971PhRvD...3.2546P}
L.~Parker,
``Quantized Fields and Particle Creation in Expanding Universes. II,''
Phys. Rev. D \textbf{3} (1971), 2546–2546


\bibitem{Chen:2016caa}
C.-M.~Chen, S.~P.~Kim, J.-R.~Sun and F.-Y.~Tang,
``Pair Production in Near Extremal Kerr-Newman Black Holes,''
Phys. Rev. D \textbf{95} (2017), 044043
[arXiv:1607.02610 [hep-th]].

\bibitem{Volovik:2022cqk}
G.~E.~Volovik,
``Particle creation: Schwinger + Unruh + Hawking,''
Pisma Zh. Eksp. Teor. Fiz. \textbf{116} (2022), 577-578
[arXiv:2206.02799 [gr-qc]].

\bibitem{Deser:1998bb}
S.~Deser and O.~Levin,
``Equivalence of Hawking and Unruh temperatures through flat space embeddings,''
Class. Quant. Grav. \textbf{15} (1998), L85-L87
[arXiv:hep-th/9806223 [hep-th]].


\bibitem{Frob:2014zka}
M.~B.~Fr{\"o}b, J.~Garriga, S.~Kanno, M.~Sasaki, J.~Soda, T.~Tanaka and A.~Vilenkin,
``Schwinger effect in de Sitter space,''
JCAP \textbf{04} (2014), 009
[arXiv:1401.4137 [hep-th]].

\bibitem{Garriga:1993fh}
J.~Garriga,
``Nucleation rates in flat and curved space,''
Phys. Rev. D \textbf{49} (1994), 6327-6342
[arXiv:hep-ph/9308280 [hep-ph]].


\bibitem{Markushevich}
A.~I.~Markushevich,
``Theory of Functions of a Complex Variable,'' Vol I and II
(Chelsea Publisher, New York, 1985).


\bibitem{Kim:2016nyz}
S.~P.~Kim,
``In-Out Formalism for One-Loop Effective Actions in QED and Gravity,''
Russ. Phys. J. \textbf{59} (2017), 1739-1745
[arXiv:1607.08719 [hep-th]].

\bibitem{Cohen:2008wz}
T.~D.~Cohen and D.~A.~McGady,
``The Schwinger mechanism revisited,''
Phys. Rev. D \textbf{78} (2008), 036008
[arXiv:0807.1117 [hep-ph]].








\end{thebibliography}
\end{document}